%% file: manuscript_final.tex
\documentclass[preprint,12pt]{elsarticle}
\usepackage{tabularx}
\usepackage[utf8]{inputenc}
\usepackage{textgreek}
\usepackage[table]{xcolor}
\newcolumntype{Y}[1]{>{\hsize=#1\hsize\linewidth=\hsize\raggedright\arraybackslash}X}
\newcolumntype{C}[1]{>{\hsize=#1\hsize\linewidth=\hsize\centering\arraybackslash}X}

\usepackage{amsmath}
\usepackage{amssymb}

\usepackage{booktabs}        
\usepackage{longtable}       
\usepackage{array}           
\usepackage{multirow}
\usepackage{makecell}        
\usepackage{threeparttable}
\usepackage{siunitx}         
\usepackage{diagbox}
\usepackage{adjustbox}
\usepackage{arydshln}
\usepackage{graphicx}
\usepackage{float}
\usepackage{caption}
\usepackage{subcaption}      
\usepackage{pdflscape}       

\usepackage{xcolor}
\definecolor{skyblue}{RGB}{0,176,240}

\biboptions{sort}  

\usepackage[
  colorlinks=true,
  linkcolor=skyblue,
  citecolor=skyblue,
  urlcolor=skyblue,
  pdfencoding=auto]{hyperref}
  
\hypersetup{
  pdftitle  = {Enhancing Resolution and Reliability with Attention Mechanisms in Nuclear Accident Surrogate Modeling},
  pdfauthor = {Wonung Jeong, Sangam Khanal, Jeongik Lee, Sangseung Lee, Joongoo Jeon}
}

\makeatletter
\def\els@aparagraph[#1]#2{\elsparagraph[#1]{#2}}
\def\els@bparagraph#1{\elsparagraph*{#1}}
\makeatother

\journal{Reliability Engineering \& System Safety}

\begin{document}

\begin{frontmatter}

\title{Enhancing Resolution and Reliability with Attention Mechanisms in Nuclear Accident Surrogate Modeling}

\author[1]{Wonung Jeong\fnref{equalcontrib}}
\author[2]{Sangam Khanal\fnref{equalcontrib}}
\author[3]{Jeongik Lee}
\author[4]{Sangseung Lee\corref{cor1}}
\ead{sangseunglee@inha.ac.kr}
\author[5]{Joongoo Jeon\corref{cor1}}
\ead{jgjeon41@postech.ac.kr}
\fntext[equalcontrib]{These authors contributed equally to this work.}
\cortext[cor1]{Corresponding authors}
\affiliation[1]{organization={Department of Quantum System Engineering, Jeonbuk National University},
            addressline={567 Baekje-daero},
            city={Jeonju},
            postcode={54896},
            country={Republic of Korea}}
\affiliation[2]{organization={Graduate School of Integrated Energy-AI, Jeonbuk National University},
            addressline={567 Baekje-daero},
            city={Jeonju},
            postcode={54896},
            country={Republic of Korea}}
\affiliation[3]{organization={Department of Nuclear and Quantum Engineering, Korea Advanced Institute of Science and Technology (KAIST)},
            addressline={291 Daehak-ro, Yuseong-gu},
            city={Daejeon},
            postcode={34141},
            country={Republic of Korea}}
\affiliation[4]{organization={Department of Mechanical Engineering, Inha University},
            addressline={100 Inha-ro, Michuhol-gu},
            city={Incheon},
            postcode={22212},
            country={Republic of Korea}}
\affiliation[5]{organization={Division of Advanced Nuclear Engineering, Pohang University of Science and Technology (POSTECH)},
            addressline={77 Cheongam-ro, Nam-gu},
            city={Pohang},
            postcode={37673},
            country={Republic of Korea}}

\begin{abstract}
Severe accident (SA) analysis of nuclear power plants relies on system codes such as MAAP, whose computational cost limits large-scale applications including dynamic probabilistic safety assessment and reinforcement learning. This study proposes the Accident Binary--Continuous Transformer (ABC-Transformer), a decoder-only surrogate combining causal self-attention with temporal attention pooling. It jointly embeds continuous thermal-hydraulic variables with binary safety-system and severe accident management guideline (SAMG) states, enabling 5-minute-resolution forecasting over 72-hour transients. Trained on 11,000 MAAP simulations of a total loss of component cooling water scenario for the OPR1000, it achieved lower one-step teacher-forcing error than CNN and recurrent baselines at fine resolutions. Autoregressive reliability was characterized as a full per-scenario distribution over 1100 held-out scenarios. For scheduled-sampling (SS) training, the median $R^2$ is $0.947 \pm 0.006$ and $3.8 \pm 1.0\%$ of scenarios fail ($R^2<0$; mean $\pm$ SD over four training seeds), while errors accumulate with rollout length. A Mahalanobis self-distance computed from each surrogate rollout, before comparison with the MAAP reference, ranks these failures (ROC-AUC $0.895$, reference seed). A threshold at the 40th percentile of calibration scores accepts $36.8 \pm 0.5\%$ of 550 evaluation scenarios, with no accepted rollout below $R^2 = 0$ for any seed, and routes the rest to MAAP.
\end{abstract}



\end{frontmatter}

\section{Introduction}
\label{sec1}

Ensuring the operational integrity and safety of nuclear power plants (NPPs) remains a central concern in nuclear engineering, as the severe accidents at Three Mile Island (TMI), Chernobyl, and Fukushima Daiichi have shown \citep{NRC1980TMI, IAEA1992INSAG7, IAEA2015Fukushima}. A severe accident is characterized by extensive reactor core degradation that, in the worst cases, can progress to containment failure and the consequent release of substantial quantities of radioactive material to the environment \citep{Sehgal2012SeverePhenomena}. The prevention of such accidents, together with effective accident response and mitigation, thus remains a major challenge in nuclear safety \citep{IAEA2019SSG54}.

A key lesson from past severe accident (SA) events is the significant role of operator actions during crisis management \citep{IAEA2015Fukushima}. Consequently, the nuclear industry has adopted measures to minimize human intervention errors, including updated emergency operating procedures (EOPs) and severe accident management guidelines (SAMGs) \citep{IAEA2019SSG54}. Nevertheless, managing an SA in real time remains a complex cognitive challenge. Operators in the main control room (MCR) face rapidly fluctuating parameters and cascading alarms under stress and time constraints \citep{Saghafi2016}. As demonstrated at TMI-2, a single misjudgment or delayed response can quickly worsen the accident \citep{NRC1980TMI}. This burden has driven a long-standing demand for analytical tools that can rapidly characterize accident behavior, both for offline preparation (SAMG validation, operator training) and, more recently, for real-time decision support \citep{Fu2023, Xiao2025RESS, Kim2026SAMG, Roma2026TrustworthyDBN}.

Among the most established of such tools are the deterministic thermal-hydraulic (TH) system codes, in particular MELCOR and the Modular Accident Analysis Program (MAAP). The nuclear engineering community has long relied on these codes to simulate plant behavior, understand accident progression, and develop mitigation strategies \citep{Sehgal2012SeverePhenomena, Cho2022Exhaustive}. These codes provide comprehensive multiphysics treatment and are designed with computational efficiency as an explicit goal. However, even this efficiency falls short of what several emerging applications require. For instance, dynamic probabilistic safety assessment seeks to evaluate tens of thousands of dynamically evolving accident trajectories rather than a handful of pre-specified sequences. At this scale, the cost of each system-code run dominates the analysis \citep{ParkLee2022DPSA, Zheng2022MultiFidelity, Jo2026DRAGON}. Likewise, real-time operator decision support \citep{LiC2024} and reinforcement-learning-based autonomous control require predictions far faster than a system-code run. This gap between the run time of system codes and the speed these applications need is the bottleneck addressed in this study.

To work around this bottleneck, recent studies have trained deep learning surrogate models on high-dimensional plant data to learn accident progression. Table~\ref{tab:nuclear_dl_papers} summarizes representative studies in this area. Most of them use recurrent neural network (RNN)-based architectures, such as long short-term memory (LSTM), gated recurrent unit (GRU), and convolutional neural network (CNN)–LSTM, to forecast critical plant parameters under specific accident scenarios \citep{Fu2023, Joo2025a, Tan2024, She2021}. However, most of these works are confined to short prediction horizons of several hundred to a few thousand seconds. The few attempts at multi-hour forecasting (e.g., 72 h), including the prior study on this dataset~\citep{Joo2025a, Lee2024RollingWindow}, have relied on coarser temporal resolutions (15 min to 1 h). This limits their ability to capture the rapid transient phenomena critical to severe accident management. More recently, Transformer-based architectures have been introduced to overcome the well-documented limitations of sequential RNNs, in particular the vanishing-gradient problem that hampers the modeling of long-range dependencies~\citep{Bengio1994Vanishing, Hochreiter1997LSTM}. They capture such dependencies through self-attention, without recurrence \citep{Vaswani2017Attention}. In the broader time-series forecasting community, efficient attention mechanisms such as Informer~\cite{zhou2021informer} and decomposition-based designs such as Autoformer~\cite{wu2021autoformer} showed that Transformers can be used for long-horizon prediction. Patch-based representations~\cite{nie2023patchtst} and inverted-attention schemes~\cite{liu2024itransformer} later became strong baselines. At the same time, simple linear models~\cite{zeng2023dlinear} have questioned whether attention is needed at all, so the choice of architecture is still open. In the nuclear domain, these methods have begun to be applied to accident parameter prediction~\cite{Tohver2023, TuoLiu2024, Zhang2025}, with reported gains in long-range dependency modeling and interpretability. To our knowledge, however, no study has yet combined long-term forecasting, fine temporal resolution, and explicit treatment of severe accident dynamics. For reliability engineering, it also matters whether a data-driven model can be trusted, that is, how confident its predictions are and whether an input lies beyond the training distribution. These questions are actively studied in the reliability community \citep{Chen2024MLReliability, Zhou2023Trustworthy, Tohme2023ReliableNN, Han2022OOD, YangKim2020Untrained}. Severe-accident surrogates, however, are still evaluated mainly with aggregate accuracy metrics, and it is rarely shown for which scenarios their predictions remain valid.

To address these gaps, this study develops the Accident Binary--Continuous Transformer (ABC-Transformer), a decoder-only Transformer surrogate of the MAAP code. From a reliability-engineering standpoint, the main aim is to determine \emph{when} the surrogate's autoregressive (AR) predictions can be trusted. The scheduled-sampling (SS)-trained model~\citep{Bengio2015ScheduledSampling} provides the principal AR results, and the archived teacher-forcing (TF)-trained model is retained for the one-step benchmark and the attention analysis. Transformer- and Temporal Fusion Transformer (TFT)-based surrogates already exist in the nuclear domain (Table~\ref{tab:nuclear_dl_papers}), and the attention components used here are not new as such. The ABC-Transformer combines them in an architecture tailored to severe accident forecasting. Causal self-attention relates the time steps of the observation window, a temporal attention pooling head aggregates the whole window for the next-step prediction, and the binary safety-system and SAMG states enter the decoder together with the continuous TH variables. This design was adopted because it improved AR prediction accuracy: in a controlled ablation, replacing the pooling head by the standard last-state readout lowered the median per-scenario $R^2$ of the AR rollout from 0.919 to 0.461 (Supplementary Material, Section S13). The attention weights also make the analysis of Section~\ref{subsec4-5} possible. The main contributions are as follows:

\begin{scriptsize}
\setlength{\tabcolsep}{4pt}
\renewcommand{\arraystretch}{1.4}
\setlength{\dashlinedash}{0.4pt}
\setlength{\dashlinegap}{1.8pt}
\begin{table}[!htbp]
\centering
\caption{Summary of recent deep-learning studies on accident parameter prediction in nuclear power plants.}
\label{tab:nuclear_dl_papers}
\setlength{\tabcolsep}{5pt}
\renewcommand{\arraystretch}{1.4}
\setlength{\dashlinedash}{0.4pt}
\setlength{\dashlinegap}{1.8pt}
\begin{adjustbox}{max width=\linewidth,center}
\begin{tabular}{@{}
  >{\centering\arraybackslash}m{3.0cm}
  >{\centering\arraybackslash}m{0.9cm}
  >{\centering\arraybackslash}m{2.6cm}
  >{\centering\arraybackslash}m{2.4cm}
  >{\centering\arraybackslash}m{2.8cm}
  >{\centering\arraybackslash}m{2.2cm}@{}}
\toprule
\textbf{Reference} & \textbf{Year} & \textbf{Model} & \textbf{Accident type} & \textbf{Inference mode} & \textbf{Prediction step} \\
\midrule
Bae et al.~\citep{Bae2026}       & 2026 & CENT                                & LOOP                          & Latent one-shot     & 1725 steps          \\
Fu et al.~\citep{Fu2023}         & 2023 & GRU (GRUS)                          & MSLB, LOCA                    & 1-step              & Not stated          \\
Joo et al.~\citep{Joo2025a}      & 2025 & CNN, LSTM, GRU, CNN--LSTM, CNN--GRU & TLOCCW                        & 1-step (AR rollout) & 72 / 144 / 288 steps \\
Tan et al.~\citep{Tan2024}       & 2024 & LSTM, RNN                           & MSLB (AP1000 PCCS)            & 1-step              & Not stated          \\
Tohver et al.~\citep{Tohver2023} & 2023 & TFT                                 & Multiple (SBLOCA, SGTR, etc.) & Multi-step          & 15 steps            \\
Tuo and Liu~\citep{TuoLiu2024}   & 2024 & Transformer                         & Shutdown transient            & 1-step              & Not stated          \\
Zhang et al.~\citep{Zhang2025}   & 2025 & EXP-Transformer                     & Stator cooling water loss     & 1-step              & Not stated          \\
\bottomrule
\end{tabular}
\end{adjustbox}
\par\smallskip
\parbox{\linewidth}{\scriptsize\textit{Abbreviations:} 
AR, autoregressive; 
CENT, consequence estimation with neural-network uncertainty; 
CNN, convolutional neural network; 
GRU, gated recurrent unit; 
LOCA, loss-of-coolant accident; 
LOOP, loss of off-site power; 
LSTM, long short-term memory; 
MSLB, main steam-line break; 
PCCS, passive containment cooling system; 
RNN, recurrent neural network; 
SBLOCA, small-break LOCA; 
SGTR, steam generator tube rupture; 
TFT, Temporal Fusion Transformer; 
TLOCCW, total loss of component cooling water.}
\end{table}
\end{scriptsize}

\begin{enumerate}

    \item \textbf{Reliability Characterization and a\newline Distance-to-Training Screen:}
    Over all 1100 test scenarios, we report the full per-scenario reliability distribution, quantify AR error accumulation (exposure bias), and show that per-scenario error is rank-correlated with a distance to the training distribution computed from the surrogate's own rollout. This distance serves as an observational reliability screen that flags scenarios for a full-code run. A screening rule at the 40th percentile of calibration self-distances (q40 rule) is evaluated on a 550-scenario evaluation half for four training seeds (Sections~\ref{sec:framework} and~\ref{sec4}).

    \item \textbf{Long-Term, High-Resolution Forecasting\newline with a Hybrid Continuous--Binary Input Design:}
    Unlike prior surrogates confined to short horizons or coarse resolutions, the ABC-Transformer forecasts severe accident progression up to 72~hours at a 5-minute resolution. A hybrid input design feeds continuous TH variables together with binary safety system actuation signals and mitigation actions directly into the decoder, providing a learning environment suited to operator-support agents.

    \item \textbf{Practical Applicability via MCR-Observable Inputs:}
    The model is trained solely on parameters observable from the MCR, and we analyze its behavior across four temporal resolutions and three lookback lengths under AR rollout.

    \item \textbf{Interpretability for Trustworthy AI:}
    To address the ``black-box'' limitation of deep surrogates, we analyze self-attention and temporal pooling weights \citep{Vaswani2017Attention, Lim2021TFT}, showing that the temporal pooling weights tend to follow physically meaningful events (such as abrupt TH changes), whereas self-attention acts mainly as an event-adaptive recency kernel.
\end{enumerate}

The remainder of this paper is organized as follows. Section~\ref{sec2} describes the severe accident dataset, including the targeted accident scenario, input variable composition, and preprocessing procedures. Section~\ref{sec3} presents the surrogate model: the forecasting task, the ABC-Transformer architecture, and its training by teacher forcing and scheduled sampling. Section~\ref{sec:framework} defines the reliability evaluation framework: accuracy metrics and failure criteria, the distance-to-training screen, and the operating point of the hybrid surrogate--MAAP strategy. Section~\ref{sec4} reports and discusses the results, covering one-step accuracy, AR reliability and exposure bias, screening performance, and attention-based interpretability. Finally, Section~\ref{sec5} summarizes the conclusions, states the limitations on how the model can be used, and outlines directions for future research.

\section{Severe Accident Dataset}
\label{sec2}\label{subsec2-1}

The SA dataset used in this study was constructed and analyzed in a previous study \citep{Lee2024RollingWindow} based on MAAP simulations for an OPR1000 NPP. The present study uses this dataset to compare different deep learning models.

The dataset comprises 11{,}000 MAAP simulations of the total loss of component cooling water (TLOCCW) scenario, each spanning the full 72-hour transient. Scenario-to-scenario variation arises from the combinations of safety-system availability and the timing of operator-initiated SAMG actions encoded by the binary inputs defined below. The full scenario-generation and sampling design is provided in the originating study \citep{Lee2024RollingWindow}. The scenarios are partitioned into 1100 held-out test scenarios and 9900 development scenarios. Rolling windows of the development scenarios are split at random into 90\% training and 10\% internal validation windows, so the internal validation is not scenario-disjoint. All reliability statistics reported in Section~\ref{sec4} are computed on the 1100-scenario test partition. The dataset covers a single initiating-event family (TLOCCW), a single plant (OPR1000), and a single code (MAAP).

\begin{figure}[tbp]
  \centering
  \makebox[\textwidth][c]{\includegraphics[width=540pt,height=0.80\textheight,keepaspectratio]{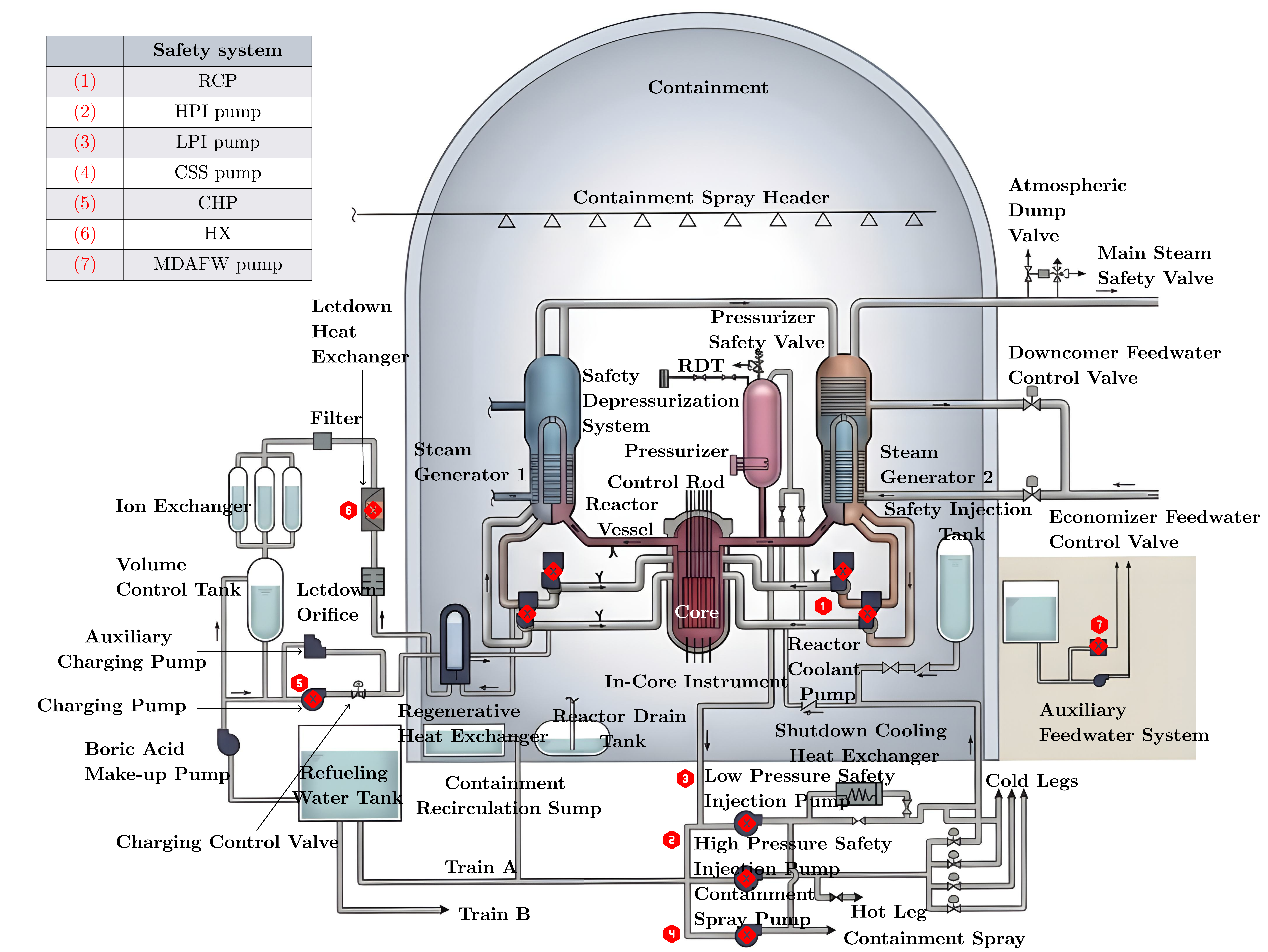}}
  \caption{Schematic illustration of possible component failures in the OPR1000 system.}
  \label{fig:opr1000_tloccw}
\end{figure}

\begin{table}[tbp]
\centering
\caption{Binary safety system and mitigation strategy variables used in this study.}
\label{tab:binary_variables}
\renewcommand{\arraystretch}{1.1}
\begin{tabular}{c l l}
\toprule
Category & Acronym & Full name \\
\midrule
~ & RCP   & Reactor coolant pump \\
~ & HPI   & High-pressure injection pump \\
~ & LPI   & Low-pressure injection pump \\
~ & CSS   & Containment spray system pump \\
~ & CHP   & Charging pump \\
~ & HX    & Heat exchanger \\
\multirow{-7}{*}{Safety system} & MDAFW & Motor-driven auxiliary feedwater pump \\
\midrule
~ & SAMG-1 & SG external injection \\
~ & SAMG-2 & RCS depressurization \\
\multirow{-3}{*}{Mitigation strategy} & SAMG-3 & RCS external injection \\
\bottomrule
\end{tabular}
\end{table}

In the TLOCCW accident, the loss of the component cooling water system adversely affects multiple safety-related and auxiliary systems essential for core cooling, and can rapidly degrade the core heat removal capability. This accident scenario was selected because a Level~2 probabilistic safety assessment (PSA) reported a high plant damage frequency for the TLOCCW event.

Under the TLOCCW accident scenario, seven safety system components, listed in Table~\ref{tab:binary_variables} and illustrated in Figure~\ref{fig:opr1000_tloccw}, were defined as binary safety system variables. In addition, three SAMG strategies were incorporated as binary inputs representing operator-initiated mitigation actions: steam generator (SG) external injection (SAMG-1), reactor coolant system (RCS) depressurization (SAMG-2), and RCS external injection (SAMG-3). Following these component failures, accident progression involving a subsequent reactor coolant pump (RCP) seal loss-of-coolant accident (LOCA), caused by seal degradation after pump trip, was further considered. Such multiple component failures and the operator-initiated SAMG interventions interact during accident progression and result in complex TH behavior.

This complex interaction can be described as a cyclic Trigger--Action--Response structure, consistent with the failure- and mitigation-driven TH values analyzed in \citep{Joo2025b}. Anomalous indications in continuous TH variables (\textit{Trigger}) prompt the activation of safety systems or mitigation strategies (\textit{Action}), whose effects are in turn reflected in subsequent TH variable changes (\textit{Response}). For this reason, both the continuous TH variables and the binary safety system/SAMG states are used as model inputs, since neither stream alone describes the accident progression completely. The binary channels encode realized actuation times from the MAAP event log. They record a fixed RCP stop; sampled heat-exchanger failure, motor-driven auxiliary feedwater pump (MDAFW) stop and SAMG initiation (pressure-gated for SAMG-1 and SAMG-3); and setpoint-triggered safety injection, charging and containment spray.

\begin{table}[tbp]
\centering
\caption{Continuous TH variables used in this study.}
\label{tab:th_variables}
\renewcommand{\arraystretch}{1.35}
\setlength{\tabcolsep}{10pt}
\begin{tabular}{c l l}
\toprule
No. & MAAP variable name & TH variable \\
\midrule
1  & PPS         & Primary system pressure        \\
2  & TGRCS(10)   & Cold leg temperature           \\
3  & TGRCS(15)   & Hot leg temperature            \\
4  & ZWV         & Reactor vessel water level     \\
5  & PSGGEN(1)   & Steam generator pressure       \\
6  & ZWDC2SG(1)S & Steam generator water level    \\
7  & MAX\_CET    & Maximum core exit temperature  \\
8  & CTMTP       & Containment pressure           \\
9  & PZRP        & Pressurizer pressure           \\
10 & PZRWL       & Pressurizer water level        \\
\bottomrule
\end{tabular}
\end{table}

The continuous TH variables considered in this study are listed in Table~\ref{tab:th_variables}. For applicability to real NPP operations, these variables were selected through quantitative analyses using the MAAP code, focusing on parameters directly observable from the MCR. Based on the Trigger--Action--Response structure described above, the proposed model was designed to forecast the evolution of TH variables during accident progression. Its inputs are the initial TH variables and the corresponding binary safety system and SAMG states.

The ten continuous TH variables span heterogeneous physical scales that differ by several orders of magnitude: pressures (MPa), temperatures (K), and levels (m). If such raw signals were fed directly to a neural network, the variables with the largest numerical range would dominate the gradient and the squared-error loss and suppress the contribution of smaller-range but physically important channels. Following standard practice in deep-learning time-series surrogates and consistent with the prior study that constructed this dataset \citep{Joo2025a, Lee2024RollingWindow}, each continuous variable was therefore independently scaled to the unit interval by min--max normalization,
\begin{equation}
\label{eq:normalization}
x' = \frac{x - x_{\min}}{x_{\max} - x_{\min}},
\end{equation}
where $x_{\min}$ and $x_{\max}$ are the per-variable minimum and maximum computed over the full dataset of 11,000 scenarios. Unless a unit is given, every error reported in this work is expressed in this normalized space, so a normalized error of $x$ corresponds to a fraction $x$ of the variable's operating range. The binary safety-system and SAMG indicators require no scaling and are used as $\{0,1\}$ conditioning inputs.

\begin{figure}[tbp]
    \centering
    \makebox[\textwidth][c]{\includegraphics[width=540pt,height=0.80\textheight,keepaspectratio]{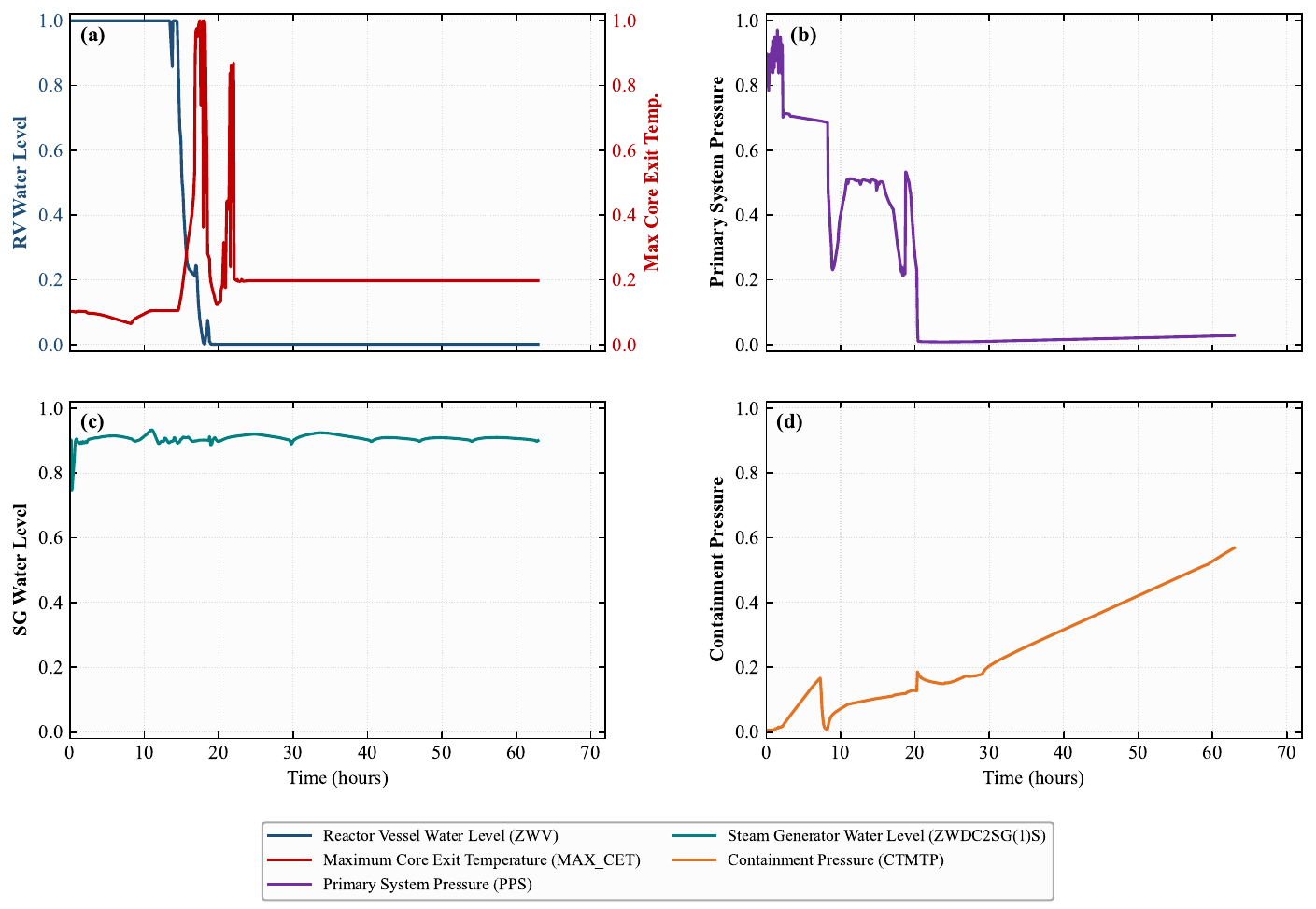}}
    \caption{Representative TH responses during a TLOCCW accident scenario over a 72-hour transient: 
    (a) reactor vessel water level (ZWV) and maximum core exit temperature (MAX\_CET); 
    (b) primary system pressure (PPS); 
    (c) steam generator water level (ZWDC2SG(1)S); and 
    (d) containment pressure (CTMTP).}
    \label{fig:th_responses_tloccw}
\end{figure}

Figure~\ref{fig:th_responses_tloccw} presents a representative 72-hour TH response from the dataset, illustrating the highly nonlinear, multi-stage progression characteristic of TLOCCW accidents. The reactor vessel water level (ZWV) in panel~(a) is nearly constant initially but drops sharply around 15--20~hours, coinciding with abrupt excursions of the maximum core exit temperature (MAX\_CET) that mark core uncovery and rapid heat-up. The primary system pressure (PPS) in panel~(b) undergoes a complex multi-step depressurization driven by the sequential activation and failure of safety injection systems. The steam generator water level (ZWDC2SG(1)S) in panel~(c) remains comparatively stable under continued auxiliary feedwater operation. The containment pressure (CTMTP) in panel~(d) rises overall, punctuated by discrete jumps and drops at key events such as RCP seal LOCA inception. This mix of slowly evolving trends and abrupt, event-driven transitions is what makes the 72-hour horizon difficult for a data-driven surrogate.

\section{Surrogate Model: ABC-Transformer}
\label{sec3}

\subsection{Forecasting Task Setup}
\label{subsec3-1}

\begin{figure}[tbp]
    \centering
    \makebox[\textwidth][c]{\includegraphics[width=540pt,height=0.80\textheight,keepaspectratio]{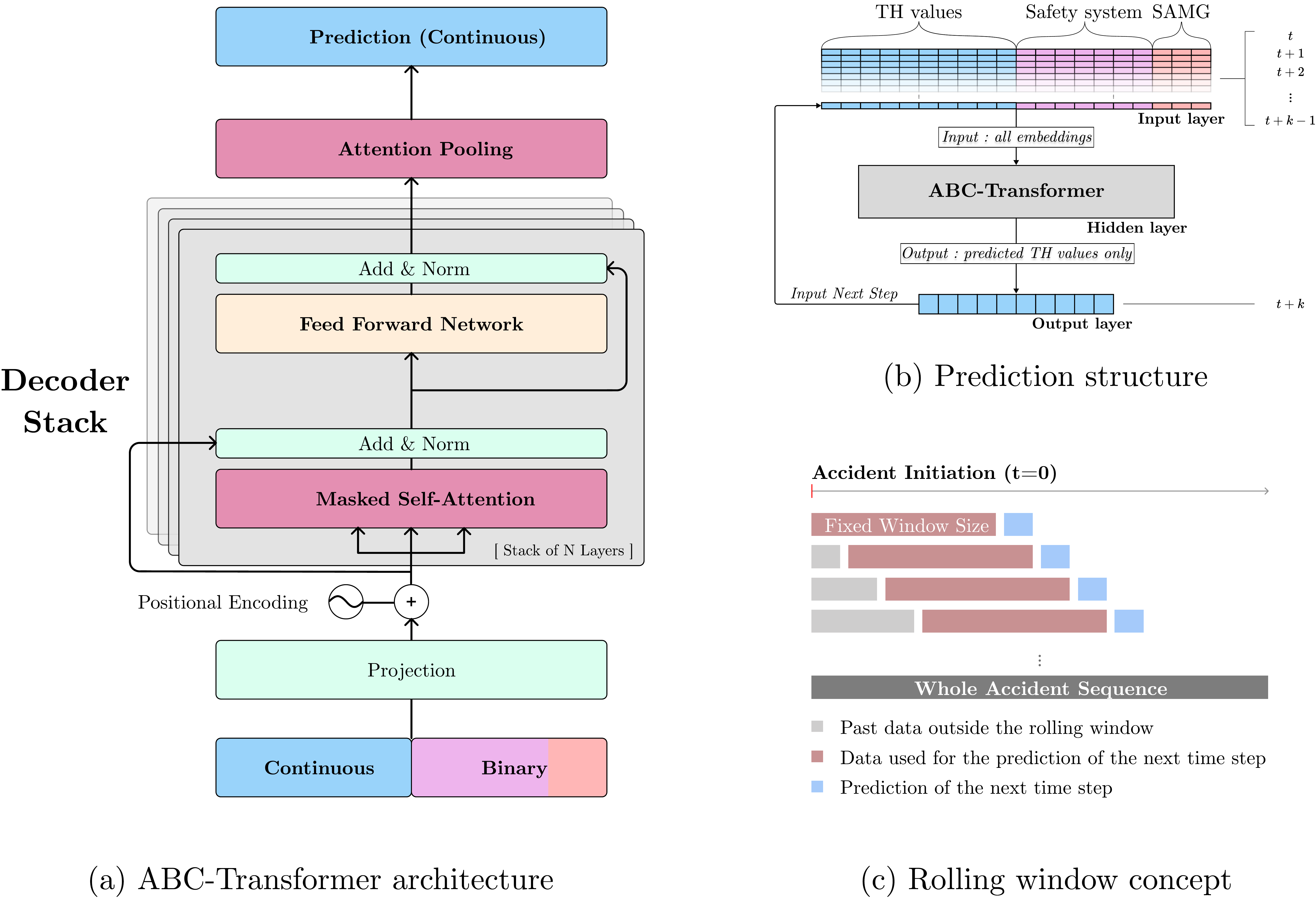}}
    \caption{Overall methodology: (a) ABC-Transformer architecture, (b) prediction structure, and (c) rolling window concept.}
    \label{fig:overall_method}
\end{figure}
The proposed ABC-Transformer is formulated as a conditional surrogate model for severe accident time-series forecasting. At each time step, the model receives both continuous TH variables and binary safety-system and SAMG states. The continuous variables are the prediction targets, whereas the binary variables are treated as scenario descriptors. Accordingly, the model forecasts the future continuous TH response conditioned on the realized actuation timeline extracted from the MAAP event log (Section~\ref{sec2}), rather than independently predicting future safety-system or operator-action states.

We formulate the severe accident forecasting task as a conditional one-step-ahead regression problem over rolling temporal windows. Let the input vector at time step $t$ be $\mathbf{x}_t \in \mathbb{R}^{D_{\mathrm{in}}}$, where $D_{\mathrm{in}} = 20$. This input vector is composed of continuous TH variables $\mathbf{c}_t \in \mathbb{R}^{10}$ and binary indicator variables $\mathbf{b}_t \in \{0,1\}^{10}$:
\begin{equation}
\label{eq:input_vector}
\mathbf{x}_t = [\mathbf{c}_t;\;\mathbf{b}_t]
\end{equation}

Our objective is to predict the continuous feature vector for the next time step $\mathbf{y}_{t+1} \in \mathbb{R}^{D_{\mathrm{out}}}$, where $D_{\mathrm{out}} = 10$ (see Figure~\ref{fig:overall_method}(b)). The model forecasts only the continuous subset, while the binary variables serve as conditioning inputs. The next continuous state is predicted deterministically from a fixed observation window of length $k$:
\begin{equation}
\label{eq:progression}
\hat{\mathbf{y}}_{t+1} = f_\theta(\mathbf{x}_{t-k+1}, \dots, \mathbf{x}_t)
\end{equation}
where $f_\theta$ denotes the ABC-Transformer with learnable parameters $\theta$, and $k$ denotes the lookback length, i.e., the number of past time steps in the observation window ($k \in \{3,10,30\}$ in this study). This $k$ is distinct from the attention key dimension $d_k$ introduced in Section~\ref{subsubsec:attention}. Throughout this paper, $k$ refers exclusively to the lookback length. The model is trained by minimizing the mean squared error (MSE) between the predicted and ground-truth continuous variables:
\begin{equation}
\label{eq:objective}
\mathcal{L}(\theta) = \frac{1}{T} \sum_{t=1}^{T} \left\| \mathbf{y}_{t+1} - f_\theta(\mathbf{x}_{t-k+1}, \dots, \mathbf{x}_t) \right\|_2^2
\end{equation}

As illustrated in Figure~\ref{fig:overall_method}, at each prediction step the input layer receives a sequence of $k$ consecutive time steps, where each time step contains TH variables, safety system signals, and SAMG variables. The ABC-Transformer processes this input sequence through its hidden layers and outputs only the 10 continuous TH variables for the immediate next time step (the step following the $k$-step input window). The binary variables are provided as known conditioning inputs and are not predicted by the model.

For long-term prediction across the entire 72-hour accident duration, we use AR inference with the rolling window mechanism depicted in Figure~\ref{fig:overall_method}(c). Starting from accident initiation ($t = 0$), a fixed-size observation window of length $k$ slides forward by one time step at each prediction iteration. Once $\hat{\mathbf{y}}_{t+1}$ is predicted, it is concatenated with the known ground-truth binary variables ($\mathbf{b}_{t+1}$) to form the new input vector $\hat{\mathbf{x}}_{t+1} = [\hat{\mathbf{c}}_{t+1};\;\mathbf{b}_{t+1}]$. The window then advances by discarding the oldest data point and incorporating the newly predicted state. This iterative scheme follows the AR forecasting paradigm
widely adopted in probabilistic time-series models such as
DeepAR~\cite{salinas2020deepar}.
\begin{align}
\label{eq:autoregressive}
\hat{\mathbf{y}}_{t+1} &= f_\theta([\mathbf{x}_{t-k+1},\;\ldots,\;\mathbf{x}_{t}]) \nonumber \\
\hat{\mathbf{y}}_{t+2} &= f_\theta([\mathbf{x}_{t-k+2},\;\ldots,\;\mathbf{x}_{t},\;\hat{\mathbf{x}}_{t+1}]) \nonumber \\
&\;\;\vdots \nonumber \\
\hat{\mathbf{y}}_{t+n} &= f_\theta([\tilde{\mathbf{x}}_{t+n-k},\;\ldots,\;\tilde{\mathbf{x}}_{t+n-1}]),
\end{align}
where $f_\theta$ denotes the ABC-Transformer model with learnable parameters $\theta$, and the window entries are defined as:
\begin{equation}
\label{eq:ar_window_def}
\tilde{\mathbf{x}}_{s} =
\begin{cases}
\mathbf{x}_{s} & \text{if } s \le t \quad (\text{observed}), \\[2pt]
\hat{\mathbf{x}}_{s} = [\hat{\mathbf{c}}_{s};\;\mathbf{b}_{s}] & \text{if } s > t \quad (\text{predicted}).
\end{cases}
\end{equation}
For $n > k$, all entries satisfy $s > t$, so the observation window contains only model-generated states $\hat{\mathbf{x}}_{s}$. This AR approach is therefore inherently susceptible to error accumulation, as prediction errors at earlier steps propagate into subsequent predictions, and Section~\ref{sec4} examines this effect in detail.

At $\Delta t = 15$~min and $k = 10$, a complete record has 289 samples: ten initial MAAP states (0--2.25~h) and 279 forecast states covering the remaining 69.75~h. The rollout also uses the realized binary actuation timeline (Section~\ref{sec2}), including setpoint-triggered actuations recorded by MAAP, but no future continuous MAAP state. A score computed from the predicted trajectory (Section~\ref{subsec:distance_screen}) is therefore available before the rollout is evaluated against the MAAP reference.

\subsection{ABC-Transformer Architecture}
\label{subsec3-2}

The proposed ABC-Transformer is designed to process heterogeneous inputs (continuous TH data and binary safety system/SAMG signals) for AR prediction of severe accident progression. Our approach builds on the Transformer architecture \citep{Vaswani2017Attention}, which replaces recurrence with the attention mechanism; this allows parallel training and relates any two positions in a sequence directly, regardless of their distance.

We adopt a decoder-only Transformer architecture, following recent decoder-only foundation models for time-series forecasting~\cite{das2024timesfm}. These models use causal, generation-oriented attention stacks for AR sequence prediction. Unlike standard encoder--decoder frameworks, our model processes the sequence in a single stream. It uses masked self-attention to strictly enforce causality, ensuring that the prediction at time $t+1$ depends only on the history up to time $t$. Whereas standard implementations rely solely on the final hidden state for prediction, our model integrates a temporal attention pooling mechanism to dynamically aggregate information across the entire historical window. As shown in Figure~\ref{fig:overall_method}(a), the model consists of four main components: an input projection layer, a positional encoding module, a stack of $N$ causal self-attention decoder layers, and a final prediction head with attention pooling, which are described in turn below.

\subsubsection{Feature Embedding and Positional Encoding}
\label{subsubsec:embedding}

The input vector $\mathbf{x}_t \in \mathbb{R}^{D_{\mathrm{in}}}$ at each time step contains both continuous and binary features that reside in fundamentally different value spaces. To map these heterogeneous inputs into a unified representation, each time step is first projected into a latent feature space of dimension $d_{\mathrm{model}}$ using a learnable linear transformation:
\begin{equation}
\label{eq:projection}
\mathbf{h}_{t}^{(0)} = \mathbf{x}_t \mathbf{W}_{\mathrm{emb}} + \mathbf{b}_{\mathrm{emb}},
\end{equation}
where $\mathbf{W}_{\mathrm{emb}} \in \mathbb{R}^{D_{\mathrm{in}} \times d_{\mathrm{model}}}$ and $\mathbf{b}_{\mathrm{emb}} \in \mathbb{R}^{d_{\mathrm{model}}}$ are learnable parameters. Applied identically across all $k$ time steps in the observation window, this transformation yields the embedded sequence $\mathbf{H}^{(0)} \in \mathbb{R}^{k \times d_{\mathrm{model}}}$. The projection maps the low-dimensional inputs to a higher-dimensional space and places the continuous and binary signals in a common embedding.

Without positional information, the self-attention mechanism treats all positions equivalently. Fixed sinusoidal positional encodings $\mathbf{PE} \in \mathbb{R}^{k \times d_{\mathrm{model}}}$ are therefore added to the embedded sequence to preserve temporal ordering information:
\begin{equation}
\label{eq:positional}
\mathbf{Z}^{(0)} = \mathbf{H}^{(0)} + \mathbf{PE}.
\end{equation}
Each entry of $\mathbf{PE}$ is defined by sine and cosine functions of varying frequencies \citep{Vaswani2017Attention}, enabling the model to encode relative positional relationships without additional trainable parameters:
\begin{align}
\label{eq:pe_formulation}
\mathrm{PE}_{(p,\,2i)}   &= \sin\!\left(\frac{p}{10000^{\,2i/d_{\mathrm{model}}}}\right), \\
\mathrm{PE}_{(p,\,2i+1)} &= \cos\!\left(\frac{p}{10000^{\,2i/d_{\mathrm{model}}}}\right),
\end{align}
where $p \in \{1, \dots, k\}$ denotes the position within the observation window and $i \in \{0, 1, \dots, d_{\mathrm{model}}/2 - 1\}$ indexes the embedding dimension. The resulting sequence $\mathbf{Z}^{(0)}$ is then passed to the stack of causal self-attention decoder layers.

\subsubsection{Causal Self-Attention Layers}
\label{subsubsec:attention}

The core representation learning is performed by a stack of $N$ identical decoder layers. Each layer consists of two sub-layers: a Masked Multi-Head Self-Attention (MHSA) module and a position-wise Feed-Forward Network (FFN).

To enforce the AR property that the prediction at time $t$ depends only on past and current inputs $\mathbf{x}_{\leq t}$, a causal mask $\mathbf{M}$ is applied to the attention computation. For a single attention head, the masked attention operation is defined as:
\begin{equation}
\label{eq:attention}
\mathrm{Attention}(\mathbf{Q},\mathbf{K},\mathbf{V}) = \mathrm{softmax}\!\left(\frac{\mathbf{Q}\mathbf{K}^T}{\sqrt{d_k}} + \mathbf{M}\right)\mathbf{V},
\end{equation}
where $\mathbf{Q}, \mathbf{K}, \mathbf{V} \in \mathbb{R}^{k \times d_k}$ are the per-head query, key, and value matrices obtained by linearly projecting the input sequence $\mathbf{Z}^{(\ell)} \in \mathbb{R}^{k \times d_{\mathrm{model}}}$, as defined in Eq.~\eqref{eq:head}. Here, $d_k = d_{\mathrm{model}}/h$ is the per-head key dimension, $h$ is the number of attention heads, and $1/\sqrt{d_k}$ is the scaling factor. The causal mask $\mathbf{M} \in \mathbb{R}^{k \times k}$ is a strictly upper-triangular matrix with $M_{ab} = -\infty$ for $b > a$ (future positions) and $M_{ab} = 0$ otherwise, so that attention weights toward any future position vanish after the softmax operation.
To capture diverse temporal patterns, the model employs MHSA, which computes $h$ parallel attention functions with independent projection parameters:
\begin{gather}
\label{eq:multihead}
\mathrm{MultiHead}(\mathbf{Z}^{(\ell)}) = \mathrm{Concat}(\mathrm{head}_1,\;\ldots,\;\mathrm{head}_h)\,\mathbf{W}^O, \\
\mathrm{head}_i = \mathrm{Attention}(\mathbf{Z}^{(\ell)}\mathbf{W}_i^Q,\;\mathbf{Z}^{(\ell)}\mathbf{W}_i^K,\;\mathbf{Z}^{(\ell)}\mathbf{W}_i^V), \quad i = 1, \dots, h, \label{eq:head}
\end{gather}
where $\mathbf{W}_i^Q, \mathbf{W}_i^K, \mathbf{W}_i^V \in \mathbb{R}^{d_{\mathrm{model}} \times d_k}$ project the input sequence into the per-head query, key, and value subspaces, and $\mathbf{W}^O \in \mathbb{R}^{d_{\mathrm{model}} \times d_{\mathrm{model}}}$ is the output projection. Each head independently attends to different temporal relationships, allowing the model to simultaneously capture both short-range fluctuations and long-range dependencies in the accident progression.

Following the attention module, a position-wise FFN applies two linear transformations with a rectified linear unit (ReLU) activation:
\begin{equation}
\label{eq:ffn}
\mathrm{FFN}(\mathbf{z}) = \mathrm{ReLU}(\mathbf{z}\mathbf{W}_1 + \mathbf{b}_1)\mathbf{W}_2 + \mathbf{b}_2
\end{equation}
where $\mathbf{z} \in \mathbb{R}^{d_{\mathrm{model}}}$ denotes the input representation at a given time step, $\mathbf{W}_1 \in \mathbb{R}^{d_{\mathrm{model}} \times d_{\mathrm{ff}}}$, $\mathbf{W}_2 \in \mathbb{R}^{d_{\mathrm{ff}} \times d_{\mathrm{model}}}$, $\mathbf{b}_1 \in \mathbb{R}^{d_{\mathrm{ff}}}$ and $\mathbf{b}_2 \in \mathbb{R}^{d_{\mathrm{model}}}$ are learnable parameters, and $d_{\mathrm{ff}}$ is the inner dimension of the feed-forward layer. Residual connections and layer normalization are applied after each sub-layer to stabilize training:
\begin{align}
\label{eq:residual}
\mathbf{Z}' &= \mathrm{LayerNorm}\bigl(\mathbf{Z}^{(\ell)} + \mathrm{MultiHead}(\mathbf{Z}^{(\ell)})\bigr), \\
\mathbf{Z}^{(\ell+1)} &= \mathrm{LayerNorm}\bigl(\mathbf{Z}' + \mathrm{FFN}(\mathbf{Z}')\bigr), \quad \ell = 0, 1, \dots, N-1,
\end{align}

\subsubsection{Temporal Attention Pooling}
\label{subsubsec:pooling}
A distinguishing feature of the ABC-Transformer architecture is the temporal attention pooling mechanism. In conventional AR models, the prediction for the next time step is typically generated from the hidden state of the final ($k$-th) position, $\mathbf{z}_k^{(N)}$, alone. However, this approach can be suboptimal for severe accident modeling. Critical events at distant past time steps, such as the onset of core uncovery or the timing of a safety injection, may be more informative for the current prediction than the immediately preceding state.
To overcome this limitation, we introduce a global temporal pooling mechanism that computes a weighted aggregate over the entire observation window of length $k$. The mechanism uses a learnable pooling query vector $\mathbf{q}_{\mathrm{pool}} \in \mathbb{R}^{d_{\mathrm{model}}}$. An attention score $s_t$ for each time step $t$ is computed via a dot product with the final decoder layer's output sequence $\mathbf{Z}^{(N)} = (\mathbf{z}_1^{(N)},\;\ldots,\;\mathbf{z}_k^{(N)})$ and then normalized into a pooling weight $\alpha_t$: 
\begin{equation}
\label{eq:pool_score}
s_t = \mathbf{z}_{t}^{(N)} \cdot \mathbf{q}_{\mathrm{pool}}
\end{equation}
\begin{equation}
\label{eq:pool_weight}
\alpha_t = \frac{\exp(s_t)}{\sum_{j=1}^{k}\exp(s_j)}
\end{equation}
The final context vector $\mathbf{c}_{\mathrm{pool}}$ is the weighted sum of all hidden states:
\begin{equation}
\label{eq:context_vector}
\mathbf{c}_{\mathrm{pool}} = \sum_{t=1}^{k}\alpha_t\,\mathbf{z}_t^{(N)}
\end{equation}
This mechanism acts as a ``soft'' temporal selection, through which the model learns which time steps in the window are most informative for the next-step forecast. The causal self-attention maps describe pairwise dependencies within each layer, whereas the pooling weights $\boldsymbol{\alpha} = [\alpha_1,\;\alpha_2,\;\ldots,\;\alpha_k]$ show how strongly each time step in the window contributes to the representation used for the final prediction. Section~\ref{sec4} uses this property to relate the predictions to physical events. A readout ablation against the last-state readout is reported in Supplementary Material, Section S13.
\subsubsection{Prediction Head}
\label{subsubsec:prediction_head}
The pooled context vector $\mathbf{c}_{\mathrm{pool}} \in \mathbb{R}^{d_{\mathrm{model}}}$ contains the aggregated temporal information of the input sequence. It is passed through a final linear layer to generate the forecast:
\begin{equation}
\label{eq:prediction}
\hat{\mathbf{y}}_{t+1} = \mathbf{c}_{\mathrm{pool}}\,\mathbf{W}_{\mathrm{out}} + \mathbf{b}_{\mathrm{out}}
\end{equation}
where $\mathbf{W}_{\mathrm{out}} \in \mathbb{R}^{d_{\mathrm{model}} \times D_{\mathrm{out}}}$ and $\mathbf{b}_{\mathrm{out}} \in \mathbb{R}^{D_{\mathrm{out}}}$ are learnable parameters, and $D_{\mathrm{out}} = 10$ corresponds to the 10 continuous TH variables.

\subsection{Training: Teacher Forcing and Scheduled Sampling}
\label{subsec:training}
The archived TF-trained model uses teacher forcing (Section~\ref{subsubsec:teacher_forcing}), and the SS-trained models that give the principal results use scheduled sampling (Section~\ref{subsubsec:scheduled_sampling}).

\subsubsection{Teacher Forcing}
\label{subsubsec:teacher_forcing}
An AR model trained for long-term prediction is susceptible to compounding errors known as exposure bias~\citep{ranzato2016exposure, Schar2025FNARX}, where small deviations in early predictions accumulate over subsequent steps. To keep training stable, teacher forcing was employed~\citep{williams1989rtrl}. Under this regime, the model is trained using the one-step-ahead formulation of Eq.~\eqref{eq:progression}, with every entry of the observation window taken from the ground-truth sequence $\mathbf{x}$ rather than from the model's own prior predictions:
\begin{equation}
\label{eq:teacher_forcing}
\hat{\mathbf{y}}_{t+1} = f_\theta(\mathbf{x}_{t-k+1},\;\mathbf{x}_{t-k+2},\;\ldots,\;\mathbf{x}_{t}).
\end{equation}
The model thus learns from accurate state trajectories, which stabilizes training and prevents error propagation during the initial learning stages. In contrast, the AR rolling-window strategy of Eq.~\eqref{eq:autoregressive} progressively replaces the ground-truth entries with predicted states $\hat{\mathbf{x}}$ and is used only at inference, to evaluate long-horizon prediction.

The base configuration adopted throughout this work and its tuning runs are documented in \ref{app:hyperparams} (Table~\ref{tab:hp-tuning}), and a one-factor-at-a-time sensitivity analysis in Supplementary Material, Section S12.

\subsubsection{Scheduled Sampling}
\label{subsubsec:scheduled_sampling}
Scheduled sampling~\citep{Bengio2015ScheduledSampling}, also applied to Transformers~\citep{Mihaylova2019SSTransformer}, exposes the model to some of its own outputs during training. In this implementation, a forward pass on the first $k-1$ input rows predicts the continuous part of the last row. This detached prediction replaces the last continuous row with probability
\begin{equation}
\label{eq:ss_probability}
p_e = \min\{0.5,\; 0.5e/8\},
\end{equation}
where $e = 0, 1, \ldots$ is the epoch index. Earlier rows and all binary values are unchanged, so training does not run through a full AR rollout. The loss of Eq.~\eqref{eq:objective} is computed against the true next state. Validation uses pure teacher forcing, and the checkpoint with the minimum validation MSE is kept. The SS recipe also differs from that of the TF-trained model in batch size, epoch limit and early-stopping patience, so comparisons between the two concern training procedures, not scheduled sampling alone. SS training used seeds 0 (reference), 42, 1 and 2 at the headline configuration ($\Delta t = 15$~min, $k = 10$) and seed 0 alone for the twelve-configuration grid (Table~\ref{tab:ss-config}).

\section{Reliability Evaluation Framework}\label{sec:framework}

This section defines accuracy metrics and failure criteria for a surrogate rollout, a self-distance screen computed from the rollout, and an operating point that accepts the surrogate rollout or routes the scenario to MAAP.

\subsection{Accuracy Metrics and Failure Criteria}\label{subsec:metrics}

To quantitatively assess predictive performance, two complementary error metrics were adopted: the
mean absolute error (MAE) and the root-mean-square error (RMSE). Both aggregate hierarchically over
the $T_n$ prediction steps of a scenario and the $N$ test scenarios. Let $\mathbf{y}^{(n)}_t$ and $\hat{\mathbf{y}}^{(n)}_t$ denote the ground-truth and predicted state vectors at step $t$ of scenario $n$. The dataset-level errors reported throughout this work, under both teacher forcing and autoregressive rollout, are the macro-averages
\begin{equation}
\label{eq:mae}
\mathrm{MAE} = \frac{1}{N}\sum_{n=1}^{N}\frac{1}{T_n}\sum_{t=1}^{T_n}\left\lVert \hat{\mathbf{y}}^{(n)}_t - \mathbf{y}^{(n)}_t \right\rVert_1,
\end{equation}
\begin{equation}
\label{eq:rmse}
\mathrm{RMSE} = \frac{1}{N}\sum_{n=1}^{N}\frac{1}{T_n}\sum_{t=1}^{T_n}\left\lVert \hat{\mathbf{y}}^{(n)}_t - \mathbf{y}^{(n)}_t \right\rVert_2,
\end{equation}
where $\lVert\cdot\rVert_1=\tfrac{1}{D}\sum_{d=1}^{D}|\cdot|$ and $\lVert\cdot\rVert_2=\sqrt{\tfrac{1}{D}\sum_{d=1}^{D}(\cdot)^2}$ denote the channel-averaged $\ell_1$ and $\ell_2$ errors over the $D=10$ continuous variables. The per-step RMSE is thus a root-mean-square across channels, and the per-scenario RMSE is the mean of these per-step values rather than a single global root-mean-square. The reliability summary in Section~\ref{subsubsec:reliability_dist}, by contrast, reports the median of the per-scenario values $\{\mathrm{MAE}^{(n)}\}$ and $\{\mathrm{RMSE}^{(n)}\}$. Here $\mathrm{MAE}^{(n)}$ and $\mathrm{RMSE}^{(n)}$ are the inner step-averages of Eqs.~\eqref{eq:mae}--\eqref{eq:rmse} evaluated for a single scenario $n$. MAE measures the average magnitude of prediction errors without directional bias, whereas RMSE, by squaring errors before averaging, disproportionately penalizes large deviations. RMSE is therefore informative for SA modeling, where extreme transient peaks and abrupt phase transitions matter for safety evaluation.

The per-scenario coefficient of determination in Section~\ref{subsubsec:reliability_dist} is the standard $R^2$ of Eq.~\eqref{eq:r2}, taken against each channel's per-scenario temporal mean:
\begin{equation}
\label{eq:r2}
R^{2}_{(n)} \;=\; 1 - \frac{\displaystyle\sum_{t=1}^{T_n}\sum_{d=1}^{D}\bigl(y^{(n)}_{t,d}-\hat y^{(n)}_{t,d}\bigr)^2}
{\displaystyle\sum_{t=1}^{T_n}\sum_{d=1}^{D}\bigl(y^{(n)}_{t,d}-\bar y^{(n)}_{d}\bigr)^2},
\qquad
\bar y^{(n)}_{d}=\frac{1}{T_n}\sum_{t=1}^{T_n} y^{(n)}_{t,d},
\end{equation}
where $D=10$ is the number of continuous channels, $T_n$ the number of rollout steps of scenario $n$, and $\bar y^{(n)}_{d}$ the per-channel temporal mean. The macro-average $R^2$ is the unweighted mean of $R^{2}_{(n)}$ over the $n=1100$ test scenarios, and the median is taken over the same set.

A scenario is a statistical failure when $R^2_{(n)} < 0$, with indicator $F_n = \mathbf{1}\{R^2_{(n)} < 0\}$. Its summed squared error then exceeds that of predicting each channel's temporal mean. This accuracy criterion is used throughout and does not measure accident severity.
Engineering checks use the maximum core-exit temperature (CET) in kelvin. With MAAP and predicted values $\Theta_{n,t}$ and $\hat{\Theta}_{n,t}$, the peak-CET error is
\begin{equation}
\label{eq:peak_error}
\delta_{\mathrm{peak},n} = \Bigl|\max_{1 \leq t \leq T_n} \hat{\Theta}_{n,t} - \max_{1 \leq t \leq T_n} \Theta_{n,t}\Bigr|.
\end{equation}
A trajectory's crossing is its first sampled state at or above 1477~K. An engineering violation ($E_n = 1$) occurs if $\delta_{\mathrm{peak},n} > 100$~K, if both trajectories cross more than 60~min apart, or if only one crosses (a missed or spurious crossing). The peak-only indicator is $P_n = \mathbf{1}\{\delta_{\mathrm{peak},n} > 100~\mathrm{K}\}$. These are study tolerances, not validated operating limits.

Medians are reported with scenario-bootstrap 95\% intervals~\citep{efron1979bootstrap} and failure fractions with Wilson 95\% intervals~\citep{wilson1927, browncai2001} (Supplementary Material, Section S1); these intervals reflect scenario sampling only. Training-seed variation is reported separately as the mean $\pm$ sample standard deviation (SD) over four SS-trained models; unless stated otherwise, values given as mean $\pm$ SD below refer to this variation.

\subsection{Distance-to-Training Reliability Screen}\label{subsec:distance_screen}

A practically important question is whether unreliable rollouts can be flagged \emph{before} they are trusted.
We screen each rollout with the self-distance $d_{\mathrm{self}}$, the Mahalanobis distance of the surrogate's own rollout states to the training distribution, defined below.
All distances are computed on the $d = 10$ continuous TH variables only. The binary safety-system/SAMG channels are excluded because a covariance structure and percentile band are not meaningful for $\{0,1\}$ indicators. Let $\boldsymbol{\mu}$ and $\boldsymbol{\Sigma}$ denote the mean and covariance of the continuous states estimated on the 9900 development scenarios (2,871,941 states at $\Delta t = 15$~min). For a continuous state $\mathbf{z} \in \mathbb{R}^{d}$, the Mahalanobis distance to training~\citep{mahalanobis1936} is
\begin{equation}
\label{eq:mahalanobis}
D_M(\mathbf{z}) = \sqrt{ (\mathbf{z} - \boldsymbol{\mu})^{\top}\, \boldsymbol{\Sigma}^{+}\, (\mathbf{z} - \boldsymbol{\mu}) }.
\end{equation}
Here $\boldsymbol{\Sigma}^{+}$ denotes the Moore--Penrose pseudo-inverse~\citep{penrose1955} of $\boldsymbol{\Sigma}$. The Mahalanobis distance standardizes by the training variance and covariance, and therefore presupposes an approximately elliptical (Gaussian) training distribution.

The self-distance of scenario $n$ averages Eq.~\eqref{eq:mahalanobis} over the forecast:
\begin{equation}
\label{eq:self_distance}
d_{\mathrm{self},n} = \frac{1}{T_n}\sum_{t=1}^{T_n} D_M\bigl(\hat{\mathbf{y}}^{(n)}_t\bigr),
\end{equation}
where $\hat{\mathbf{y}}^{(n)}_t$ is the predicted continuous state at forecast step $t = 1, \dots, T_n$. The score is derived from the surrogate model alone: it uses the surrogate's own rollout and the frozen training statistics $\boldsymbol{\mu}$ and $\boldsymbol{\Sigma}$. It is available once the rollout is complete, before the rollout is evaluated against the MAAP reference trajectory.
A small self-distance cannot rule out a large error: an incorrect rollout can stay close to the training states, and the time average can dilute short excursions.

\subsection{Operating Point and Hybrid Surrogate--MAAP Strategy}\label{subsec:operating_point}

A fixed permutation (seed~42) of the sorted IDs splits the 1100 held-out scenarios into calibration and evaluation subsets $\mathcal{C}$ and $\mathcal{E}$ of 550 each, identical for the four SS-trained models. The threshold $\tau_q$ is the empirical $q$-quantile (linear interpolation) of the model's calibration self-distances (Eq.~\eqref{eq:self_distance}), frozen before evaluation:
\begin{equation}
\tau_q = Q_q\{d_{\mathrm{self},n} : n \in \mathcal{C}\}, \qquad A_n = \mathbf{1}\{d_{\mathrm{self},n} < \tau_q\}.
\label{eq:coverage_threshold}
\end{equation}
Scenarios with $A_n = 1$ are accepted; the others, ties included, are routed to MAAP. With $F_n = \mathbf{1}\{R^2_{(n)} < 0\}$ (Eq.~\eqref{eq:r2}), the accepted count, accepted failures, coverage and retained-set failure fraction on $\mathcal{E}$ are
\begin{equation}
n_a = \sum_{n \in \mathcal{E}} A_n, \qquad k_f = \sum_{n \in \mathcal{E}} A_n F_n, \qquad c = \frac{n_a}{550}, \qquad \hat{r}_{\mathrm{acc}} = \frac{k_f}{n_a}.
\label{eq:screen_metrics}
\end{equation}
Failure recall is the fraction of evaluation failures routed to MAAP, and $n_a^{-1}\sum_{n \in \mathcal{E}} A_n E_n$ is the accepted engineering-violation fraction (CET checks, Eq.~\eqref{eq:peak_error}).
For $n_a > 0$, the nominal one-sided 95\% Clopper--Pearson upper bound \citep{clopperpearson1934} is
\begin{equation}
U_{0.95} =
\begin{cases}
1 - 0.05^{1/n_a}, & k_f = 0,\\
\mathrm{Beta}^{-1}(0.95;\, k_f + 1,\, n_a - k_f), & 0 < k_f < n_a,\\
1, & k_f = n_a,
\end{cases}
\label{eq:cp_bound}
\end{equation}
with $\mathrm{Beta}^{-1}$ the beta quantile function. The bound is model-wise, not joint over the four models, and nominal because the operating point was developed on these scenarios.

The q40 rule sets $q = 0.40$, a target acceptance fraction on calibration data rather than a routing budget. The threshold uses only calibration self-distances, without failure labels or MAAP trajectories beyond the initial window, so $\tau$ is specific to the surrogate model. The 40\% level itself was chosen by comparing operating points on the same model and scenarios, including a learn-then-test rule and a rule routing 35\% of the calibration scenarios. Only q40 accepted zero rollouts with $R^2<0$ in all four seeds and kept a stable threshold (Section~\ref{subsubsec:q40_results}; Supplementary Material, Section S3). Because it was chosen on these data, the level is specific to this model and scenario set.

The workflow has three steps. (1) Supply the initial window and the actuation timeline. (2) Run the surrogate rollout and compute $d_{\mathrm{self}}$. (3) Accept the rollout if $d_{\mathrm{self}} < \tau$; otherwise, route the scenario to a full MAAP run. For the reference SS-trained model, $\tau = 2.2484$ (Section~\ref{subsubsec:q40_results}). The routed fraction $r = 1 - c$ is the demand for MAAP confirmation runs, not a measured saving: here the setpoint-triggered actuation times came from the MAAP event log (Section~\ref{sec2}), and the MAAP runtime was not benchmarked.
The hybrid surrogate--MAAP output, evaluated over all evaluation scenarios, keeps the surrogate trajectory if accepted and the MAAP reference if routed. With the peak-only indicator $P_n$ and $v_{\mathrm{peak,acc}} = n_a^{-1}\sum_{n \in \mathcal{E}} A_n P_n$, its peak-CET agreement is
\begin{equation}
G_{\mathrm{hybrid}} = 1 - c\, v_{\mathrm{peak,acc}},
\label{eq:hybrid_agreement}
\end{equation}
a property of the combined workflow. Training-distance thresholds q90/q92, a learn-then-test rule and the calibration-bootstrap sensitivity are reported in Supplementary Material, Section S3.

\section{Results and Discussion}
\label{sec4}

\subsection{TF Accuracy across Resolutions and Lookbacks}
\label{subsec4-1}

Table~\ref{tab:teacher_forcing_results} compares the TF-trained ABC-Transformer against five baselines from a previous study~\citep{Joo2025a} under the teacher-forcing one-step protocol. Errors are measured by MAE and RMSE~\citep{Hyndman2006Forecast} (Section~\ref{subsec:metrics}, Eqs.~\eqref{eq:mae}--\eqref{eq:rmse}). Both studies used the same dataset and the same split ratio, sampling train/test partitions randomly (Section~\ref{subsec2-1}). The sampling was independent, so the two test sets are comparable but not identical. The lookback was fixed at $k = 3$, the value used in~\citep{Joo2025a}. A separate model was trained at each resolution, as in the baselines. All errors use the normalized scale of Eq.~\eqref{eq:normalization}, and all values in this table, baseline and proposed, are single-seed point estimates.

\begin{table}[htbp]
\centering
\caption{Teacher-forcing performance comparison (MAE and RMSE) between baseline models from previous studies and the proposed ABC-Transformer. The ABC-Transformer row is the archived TF-trained model.}
\label{tab:teacher_forcing_results}
\begin{adjustbox}{max width=\textwidth, center}
\begin{threeparttable}
\renewcommand{\arraystretch}{1.2}
\begin{tabular}{lcccccccc}
\toprule
& \multicolumn{2}{c}{$\Delta t = 60$~min} & \multicolumn{2}{c}{$\Delta t = 30$~min} & \multicolumn{2}{c}{$\Delta t = 15$~min} & \multicolumn{2}{c}{$\Delta t = 5$~min} \\
\cmidrule(lr){2-3}\cmidrule(lr){4-5}\cmidrule(lr){6-7}\cmidrule(lr){8-9}
\multirow{-2}{*}{Model} & MAE & RMSE & MAE & RMSE & MAE & RMSE & MAE & RMSE \\
\midrule
CNN       & 0.0067 & 0.0170 & 0.0058 & 0.0153 & 0.0050 & 0.0140 & -- & -- \\
LSTM      & 0.0095 & 0.0349 & 0.0074 & 0.0262 & 0.0053 & 0.0186 & -- & -- \\
GRU       & 0.0106 & 0.0355 & 0.0081 & 0.0265 & 0.0064 & 0.0197 & -- & -- \\
CNN--LSTM & \textbf{0.0063} & 0.0163 & 0.0060 & 0.0149 & 0.0051 & 0.0129 & -- & -- \\
CNN--GRU  & \textbf{0.0063} & 0.0163 & 0.0054 & 0.0143 & 0.0049 & 0.0131 & -- & -- \\
ABC-Transformer & 0.0070 & \textbf{0.0116} & \textbf{0.0045} & \textbf{0.0078} & \textbf{0.0032} & \textbf{0.0058} & 0.0022 & 0.0043 \\
\bottomrule
\end{tabular}
\begin{tablenotes}[flushleft]
\small
\item \textit{Note:} Boldface indicates the best performance for each metric among models evaluated at that resolution. The $\Delta t = 5$~min column is left unbolded because the previous studies did not evaluate at this resolution.
\end{tablenotes}
\end{threeparttable}
\end{adjustbox}
\end{table}

The ABC-Transformer attains the lowest RMSE at every resolution evaluated in~\citep{Joo2025a}, and the lowest MAE at $\Delta t = 30$ and $15$~min. The margin is widest at $\Delta t = 15$~min, where the MAE is 0.0032 and the RMSE is 0.0058. These values are nominally 35.3\% and 54.8\% lower (from the unrounded values) than those of CNN--GRU (MAE 0.0049) and CNN--LSTM (RMSE 0.0129), the best baselines for each metric.

At the coarsest resolution, $\Delta t = 60$~min, the model is \emph{not} uniformly best. Its MAE of 0.0070 is 11.1\% higher than CNN--LSTM and CNN--GRU (0.0063), but its RMSE of 0.0116 is 28.8\% lower. A lower RMSE at a moderately higher MAE suggests a more uniform error distribution with fewer large excursions. This is also reflected in the RMSE/MAE ratio, which is 1.66 for the ABC-Transformer, against 2.54 to 3.67 for the baselines (Supplementary Material, Table S4.1).

The one-step advantage therefore depends on resolution, and two caveats should be kept in mind. First, the baseline values are single-run estimates from~\citep{Joo2025a}, reported without seeds or confidence intervals (CIs), so the reductions above are indicative rather than significance-tested. Second, one-step accuracy under teacher forcing says little about the autoregressive rolling-window setting in which the surrogate is used, and no published baseline reports comparable metrics for that setting. Table~\ref{tab:teacher_forcing_results} should thus be read as a sanity check on the architecture rather than as evidence of deployment-level superiority (Section~\ref{subsec4-2}).

One-step accuracy also improves monotonically with resolution (MAE $0.0070 \to 0.0022$, RMSE $0.0116 \to 0.0043$ from $\Delta t = 60$ to $5$~min). This reflects both the richer transient dynamics exposed to the attention mechanism (Section~\ref{subsec4-5}) and the arithmetically smaller, hence easier, per-step increment. The same refinement, however, multiplies the number of autoregressive steps over a fixed physical horizon. Section~\ref{subsubsec:reliability_dist} examines whether the apparent gain survives the resulting error accumulation. At the headline configuration, the SS-trained model has a one-step MAE and RMSE of $0.0030$ and $0.0054$, so SS training does not degrade one-step accuracy there.

The TF-trained ABC-Transformer was also evaluated under teacher forcing at four time resolutions ($\Delta t = 60, 30, 15, 5$~min) and three lookbacks ($k = 3, 10, 30$). One-step performance was measured by predicting the continuous TH variables at $t+1$ from ground truth up to $t$. The results (Supplementary Material, Table S5.1) are averaged over the full evaluation period.
Under teacher forcing, all three lookbacks reach a similarly low one-step error. The MAE/RMSE differences among $k=3$, $10$, and $30$ fall at the fourth decimal place at every resolution (at $\Delta t=15$~min, RMSE $0.0058$/$0.0055$/$0.0057$), within the run-to-run noise of single-seed training. These differences do not support a per-resolution optimum, so we designate no ``best'' $k$. This is expected: when every input step is ground truth, a window that covers the immediate dynamics already holds most of the information needed, so the lookback length matters little. Its effect appears under AR rollout, where errors propagate (Section~\ref{subsec4-3}).

\subsection{Autoregressive Reliability and Exposure Bias}
\label{subsec4-3}
Because the autoregressive results have no directly comparable external baseline, we describe the surrogate's reliability by its full per-scenario distribution instead of a single summary statistic. Unless stated otherwise, every analysis in this subsection uses the SS-trained model (reference seed~0) at the headline configuration ($\Delta t = 15$~min, $k = 10$) over all $n = 1100$ held-out test scenarios, with the TF-trained model as the reference comparison.

Three caveats apply to all results in this subsection. First, unless a unit is given, all errors are expressed in the normalized space of Eq.~\eqref{eq:normalization}; we make no conservative-margin or safety-qualification claim. Second, training-seed variation is assessed separately with four SS-trained models (Section~\ref{subsubsec:training_seeds}), and the confidence intervals below bound scenario-sampling uncertainty only. Third, the binary safety-system and SAMG channels are supplied as known conditioning inputs, so the task is conditional trajectory emulation given the realized actuation timeline, not autonomous prediction of operator actions. Freezing the binary inputs at their initial-window values lowers the median $R^2$ to $-4.68$ (TF-trained model; Supplementary Material, Section S6).

\subsubsection{Resolution, lookback, and configuration selection}
\label{subsec4-2}

\paragraph{Autoregressive error accumulation.}
Teacher forcing measures one-step accuracy, but a surrogate used in practice runs autoregressively, feeding its own predictions back as inputs for the following steps. Table~\ref{tab:AR} presents the AR forecasting performance of the SS-trained model averaged over all test scenarios.
\begin{table}[tbp]
\centering
\caption{Autoregressive forecasting performance (MAE / RMSE) of the SS-trained model (seed 0) on the OPR1000 test dataset across time resolutions and lookbacks, averaged over all scenarios. The TF-trained counterpart is Supplementary Table S8.2.}
\label{tab:AR}
\renewcommand{\arraystretch}{1.2}
\small
\begin{adjustbox}{max width=\linewidth,center}
\begin{tabular}{c cc cc cc}
\toprule
\diagbox{\makecell{Time\\resolution}}{Lookback}
& \multicolumn{2}{c}{$k = 3$}
& \multicolumn{2}{c}{$k = 10$}
& \multicolumn{2}{c}{$k = 30$} \\
\cmidrule(lr){2-3}\cmidrule(lr){4-5}\cmidrule(lr){6-7}
& MAE & RMSE & MAE & RMSE & MAE & RMSE \\
\midrule
$\Delta t = 60$ min & 0.0222 & 0.0391 & 0.0170 & 0.0295 & 0.0129 & 0.0225 \\
$\Delta t = 30$ min & 0.0298 & 0.0531 & 0.0189 & 0.0340 & 0.0154 & 0.0276 \\
$\Delta t = 15$ min & 0.0519 & 0.0845 & 0.0257 & 0.0468 & 0.0207 & 0.0369 \\
$\Delta t = 5$ min  & 0.0723 & 0.1163 & 0.0457 & 0.0749 & 0.0544 & 0.0852 \\
\bottomrule
\end{tabular}
\end{adjustbox}
\end{table}
Unlike teacher forcing, AR rollout shows a clear effect of resolution and lookback, since errors propagate through the rolling window. Relative to the TF-trained model, the SS recipe lowers the mean AR MAE in 11 of the 12 configurations, by 15--56\%; the exception is $\Delta t = 15$~min with $k = 3$ (+4\%; Supplementary Table S8.2). Two trends can be observed.

First, no single lookback dominates once the full per-scenario distribution is considered. For the TF-trained model (Supplementary Table S8.4), a longer lookback raises the median per-scenario $R^2$ at the finer resolutions but degrades the lower tail. Relative to $k=10$, $k=30$ raises the fraction of scenarios with $R^2 < 0$ (e.g.\ $9.7\% \to 19.8\%$ at $\Delta t = 15$~min) and lowers the 10th-percentile $R^2$. For the SS-trained model (Table~\ref{tab:t2-ksens}), $k = 30$ also gives the higher median at $\Delta t = 15$~min ($0.962$ versus $0.938$) at a similar failure fraction ($3.5\%$ versus $3.2\%$).

Second, error grows with the number of AR steps, as expected from exposure bias. A 72-hour scenario spans 73 samples at $\Delta t = 60$~min but 865 at $\Delta t = 5$~min. For the SS-trained model, the median per-scenario $R^2$ of the 5-min rollout ($k=10$) is $0.919$, compared with $0.963$ for the 60-min rollout. However, $5.8\%$ of the 5-min rollouts fall below $R^2 = 0$ (TF-trained model: $0.715$, $0.948$ and $29.4\%$; Supplementary Table S8.4). We retain $\Delta t = 5$~min because the fast transients governing severe-accident progression (core uncovery, depressurization, core-exit temperature excursions) motivate evaluating this cadence.

The headline configuration follows an explicit rule: within $\Delta t \leq 15$~min, minimize the failure fraction ($R^2<0$) first, then maximize the median $R^2$. For the TF-trained model, the $5$-min resolution attains the lowest one-step teacher-forcing error. Under rollout, however, it incurs a far larger failure fraction ($21$--$30\%$ across lookbacks, against $5$--$20\%$ at coarser resolutions; Supplementary Table S8.4). We therefore adopt the finest resolution that avoids this penalty, $\Delta t = 15$~min. At this resolution, $k = 10$ minimizes that fraction for both training recipes ($3.2\%$ SS-trained, Table~\ref{tab:t2-ksens}; $9.7\%$ TF-trained). A median-only ranking would instead prefer $k = 30$ (median $0.962$ SS-trained and $0.929$ TF-trained). We nevertheless give priority to the failure fraction, because for a safety-relevant surrogate the rare catastrophic rollouts matter more than the typical case.

\begin{figure}[tbp]
    \centering
    \makebox[\textwidth][c]{\includegraphics[width=540pt,height=0.80\textheight,keepaspectratio]{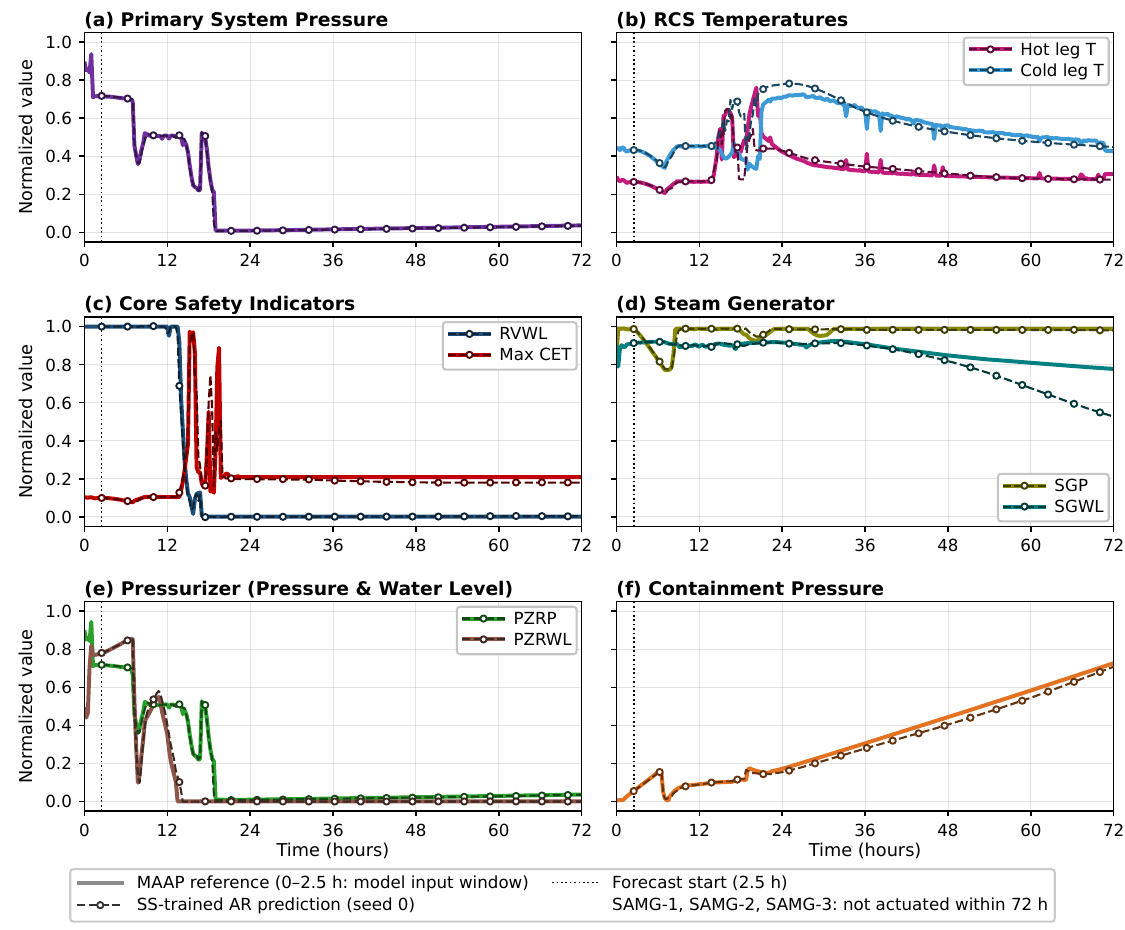}}
    \caption{Autoregressive long-term prediction of the SS-trained ABC-Transformer (seed 0) for a representative TLOCCW scenario (Scenario~10070): 
(a) primary system pressure; 
(b) RCS temperatures; 
(c) core safety indicators; 
(d) steam-generator variables; 
(e) pressurizer variables; and 
(f) containment pressure. 
Solid lines denote the MAAP reference and dashed lines with open circles the prediction. The dotted line marks the forecast start ($t = 2.5$~h); no SAMG strategy is actuated in this scenario. Configuration: $\Delta t = 15$~min, $k = 10$.}
    \label{fig:ar_prediction_scenario6736}
\end{figure}
Figure~\ref{fig:ar_prediction_scenario6736} presents a typical case (Scenario~10070; $R^2 = 0.939$, close to the median $0.938$ of all 1100 scenarios), which the q40 rule of Section~\ref{subsec:operating_point} accepts. The surrogate follows both regimes of the transient: the fast excursions of the early phase (sharp peaks in MAX\_CET and hot-leg temperature near $15$--$20$~h) and the slow drift of containment pressure over the remaining horizon. The accepted rollout is not free of error, however. After about $40$~h the predicted steam-generator water level drifts below the MAAP reference (panel~(d)), and over the four seeds $8.8 \pm 4.7\%$ of accepted rollouts violate a CET engineering criterion (Section~\ref{subsubsec:q40_results}). Section~\ref{subsubsec:reliability_dist} analyzes the complete per-scenario reliability distribution, including the worst cases.

\subsubsection{Per-scenario reliability distribution and exposure bias}
\label{subsubsec:reliability_dist}

\input{Table06_reliability_distribution}

The aggregate errors hide a strongly skewed distribution across scenarios. Table~\ref{tab:t1-reliability-dist} reports the per-scenario $R^2$ distribution at this configuration, and \ref{app:reliability} gives the full per-configuration grid. The median per-scenario $R^2$ of the SS-trained model is $0.938$, well above the macro-average of $0.836$. The gap arises because the macro-average, as an arithmetic mean, is pulled down disproportionately by a left tail of poorly predicted scenarios (in particular the $3.2\%$ below $R^2 = 0$). The median, in contrast, is insensitive to such outliers. Typical scenarios are therefore reproduced well ($R^2 \approx 0.94$), but failures are frequent enough to matter. The 10th-percentile $R^2$ is $0.652$; $3.2\%$ of all scenarios fall below $R^2 = 0$, that is, they perform worse than a per-scenario mean predictor, and $6.1\%$ fall below $R^2 = 0.5$. For a safety-relevant surrogate these tail statistics need to be reported alongside the median, because rare large failures, and not only typical accuracy, limit how the model can be used. Table~\ref{tab:t1-reliability-dist} gives the corresponding $95\%$ confidence intervals; the metrics are defined in Section~\ref{subsec:metrics}, and the interval estimation is described in Supplementary Material, Section S1. For the TF-trained model, the median, macro-average and failure fraction are $0.899$, $0.604$ and $9.7\%$ on the same basis (Supplementary Table S8.3).

\paragraph{Exposure bias and error accumulation.}
\label{subsubsec:exposure}

\begin{figure}[htb]
    \centering
    \includegraphics[width=\linewidth]{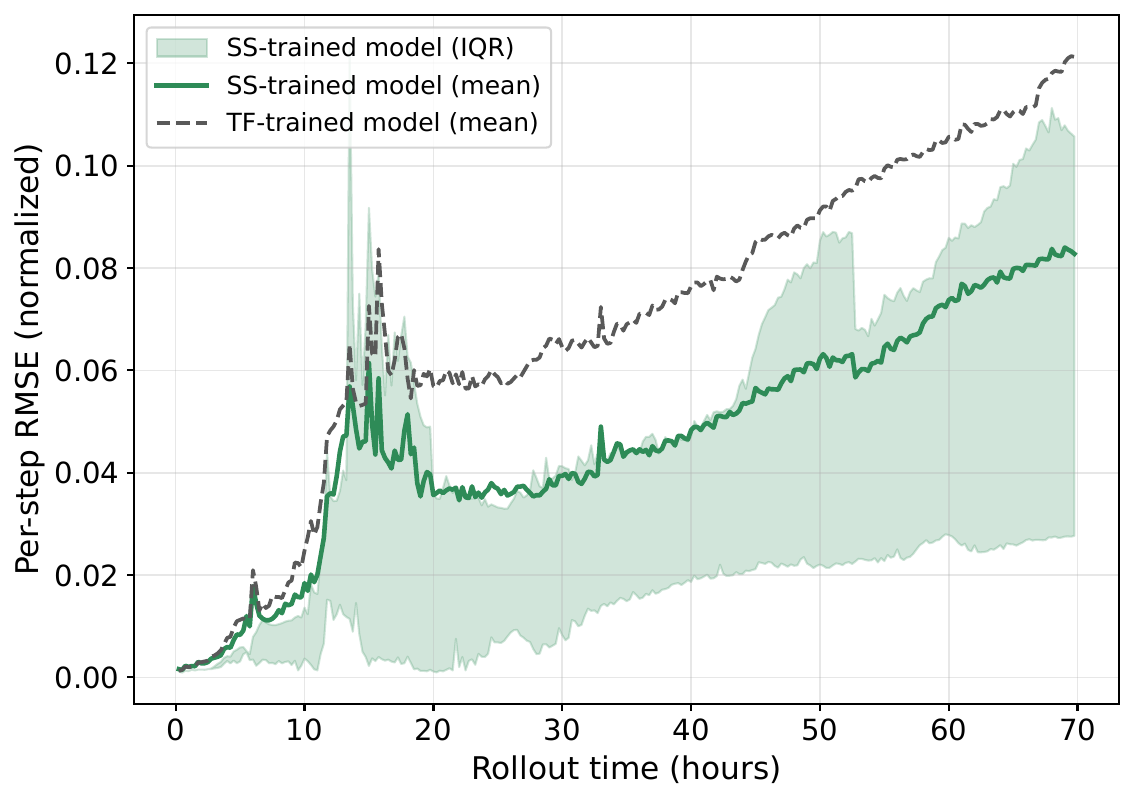}
    \caption{Autoregressive per-step RMSE versus rollout time at the headline configuration for the SS-trained model (solid, mean; shaded band, inter-scenario interquartile range) and the archived TF-trained model (dashed, mean). The scenario count decreases from 1100 to 1094 at the final step.}
    \label{fig:error_vs_step}
\end{figure}

The gap between teacher-forcing and AR accuracy is a quantitative indication of exposure bias~\citep{ranzato2016exposure}. Supplementary Table S7.1 reports the autoregressive-to-teacher-forcing (AR/TF) error ratio for every configuration of the TF-trained model. The MAE-based ratio grows monotonically with the number of rollout steps, from roughly $3\text{--}6\times$ at $\Delta t = 60$~min (73 samples) to $13\text{--}16\times$ at $\Delta t = 15$~min (289 samples) and $38\text{--}46\times$ at $\Delta t = 5$~min (865 samples). For the SS-trained model at the headline configuration, the ratio is $8.7\times$ (Supplementary Table S7.2). Part of this ratio is expected by definition, since a multi-step rollout accumulates more error than one-step teacher forcing. Two observations nevertheless suggest that errors also compound beyond this. First, even the best-predicted scenario of the TF-trained model (6736; Supplementary Table S10.1) has an AR MAE of $0.0053$, against the $0.0030$ one-step teacher-forcing error of the same configuration. Because this scenario stays close to the training data throughout (mean Mahalanobis distance $2.5$), the gap appears without a large departure from the training distribution. Second, as can be seen in Figure~\ref{fig:error_vs_step}, the per-step error grows overall along the rollout. The mean per-step RMSE of the SS-trained model rises from $0.0016$ at the first step to $0.083$ by the final step (TF-trained: $0.0013$ to $0.121$), and the inter-scenario spread widens.

\subsubsection{Training-seed variability and comparison with the TF-trained model}
\label{subsubsec:training_seeds}
Over the four training seeds, the SS-trained models reach an AR MAE of $0.0259 \pm 0.0010$, an RMSE of $0.0469 \pm 0.0015$ and a median $R^2$ of $0.947 \pm 0.006$ (Table~\ref{tab:ss-seed-summary}), with a failure fraction of $3.8 \pm 1.0\%$ (35, 53, 30 and 51 of 1100). Seed~0, used as the reference, has the lowest median of the four and was not selected for performance. These SDs describe the variation between trained models, not scenario sampling, and the twelve-configuration grid uses a single seed.

The SS-trained model of Section~\ref{subsubsec:scheduled_sampling} was evaluated on the same 1100-scenario test set. SS training lowers the average error: the AR MAE falls from $0.0413$ to $0.0257$ ($-38\%$) and the AR RMSE from $0.0697$ to $0.0468$ ($-33\%$). The teacher-forcing one-step validation error at this configuration is essentially unchanged (marginally lower), which suggests that the reduction comes from less error accumulation and not from a better one-step fit. Accordingly, the per-step error is lower over almost the entire rollout (Figure~\ref{fig:error_vs_step}; lower at $97\%$ of steps, with the final-step RMSE reduced from $0.121$ to $0.083$). On the same $R^2$ basis (Eq.~\eqref{eq:r2}), the per-scenario reliability distribution improves together with the aggregate error. The macro-average per-scenario $R^2$ rises from $0.604$ to $0.836$, the median from $0.899$ to $0.938$, and the failure fraction is more than halved, from $9.7\%$ to $3.2\%$ (full before-and-after comparison in Supplementary Table S8.1). At this configuration, SS training therefore reduces both the aggregate error and the catastrophic rollouts that dominate the reliability tail. Because the recipes also differ in batch size, epoch limit and early-stopping patience (Table~\ref{tab:ss-config}), the comparison is between training procedures and does not isolate the effect of scheduled sampling. The gain is also unlikely to be a seed artifact: the archived TF-trained model's failure fraction ($9.7\%$) is lower than that of four seeds retrained with the TF training pipeline ($11.2$--$17.2\%$; Supplementary Table S8.5). The main countervailing effect is a mild reduction in the near-perfect fraction (the proportion with $R^2 \geq 0.99$ falls from $23.0\%$ to $18.9\%$), possibly because training on self-generated inputs acts as a regularizer. The SS-trained model is therefore used for all subsequent reliability analyses, and the TF-trained model is kept for the one-step benchmark (Section~\ref{subsec4-1}) and for the attention analysis of its archived checkpoint (Section~\ref{subsec4-5}).

\subsection{Screening Performance and Failure Mitigation}\label{subsec:screening}

This section evaluates the self-distance screen (Section~\ref{subsec:distance_screen}) and the q40 rule (Section~\ref{subsec:operating_point}) for the SS-trained models.

\subsubsection{Association between distance and rollout error}
\label{subsubsec:ood_screen}

\begin{figure}[tbp]
    \centering
    \makebox[\textwidth][c]{\includegraphics[width=540pt,height=0.80\textheight,keepaspectratio]{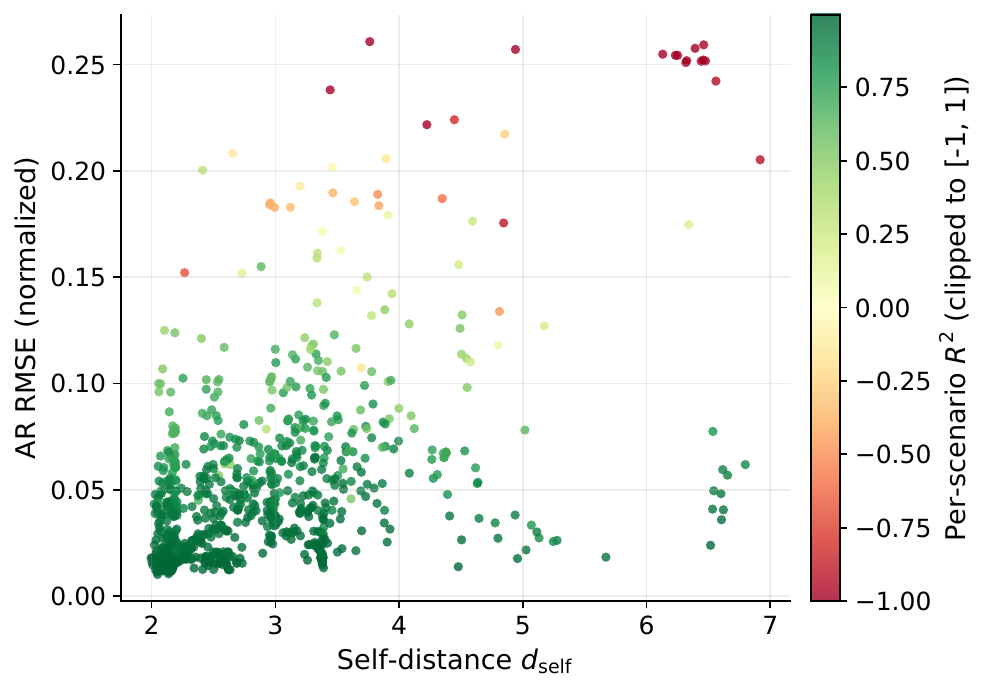}}
    \caption{Self-distance $d_{\mathrm{self}}$ (Eq.~\eqref{eq:self_distance}) versus per-scenario AR RMSE for the SS-trained model (seed 0, $n = 1100$), colored by per-scenario $R^2$ (clipped to $[-1,1]$). Only the error and the color require the MAAP reference.}
    \label{fig:ood_scatter}
\end{figure}

For the SS-trained reference model, $d_{\mathrm{self}}$ correlates with scenario RMSE at Spearman $\rho = 0.576$ (Pearson $r = 0.556$; Table~\ref{tab:t4-ood}, Fig.~\ref{fig:ood_scatter}). Its area under the receiver operating characteristic curve (ROC-AUC) for $R^2<0$ is 0.895 (35 failures; 0.872 for $R^2<0.5$); the AUC measures how well the score ranks the whole cohort and is not a failure probability. The retrospective distance $d_{\mathrm{retro},n}$ is the time mean of Eq.~\eqref{eq:mahalanobis} over the true MAAP states of the forecast window (Supplementary Material, Section S2). It reaches $\rho = 0.649$ with the SS-trained RMSE but needs the MAAP reference trajectory and serves only as a diagnostic.
The percentile-band exceedance (PBE) is the fraction of time-channel cells outside fixed, channel-specific $[1,99]$ percentile intervals of the training states (Supplementary Material, Section S2), not the per-step bands of Section~\ref{subsubsec:case_study}. On the predicted trajectory, PBE is weakly and negatively rank-correlated with the SS-trained RMSE ($\rho = -0.151$) and is not used as a screen.

\input{Table07_distance_screen}

For the TF-trained model, $d_{\mathrm{retro}}$ uses the Mahalanobis distance~\citep{mahalanobis1936} with the mean and covariance of the 9900 development scenarios. The per-scenario mean distance correlates with AR RMSE at Spearman $\rho = 0.70$~\citep{spearman1904} (95\% CI $[0.67, 0.73]$; Supplementary Table S2.1). The failed scenarios ($R^2<0$) cluster at the largest distances ($3.73$ versus $2.38$ for $R^2 \geq 0.9$). Cliff's $\delta = +0.90$ indicates a near-complete separation ($\delta = +1$ would mean every failed scenario lies farther from training than every reliable one). Computed on the true trajectory, PBE shows the same monotonic association with the TF-trained model's AR RMSE ($\rho = 0.57$; Supplementary Fig.~S2.1 and Table S2.1). As both are computed over the full population, this association is population-level, unlike the two-scenario comparison of Section~\ref{subsubsec:case_study}.

The association holds in three robustness checks (Supplementary Material, Section S2). It is stable across folds and all six resolution/lookback configurations ($\rho \in [0.68, 0.73]$). A partial correlation controlling for accident-severity proxies remains significant ($\rho = 0.55$), so the association is not explained by these proxies alone. A nonparametric $k$-nearest-neighbor distance~\citep{sun2022knn} correlates even more strongly ($\rho = 0.79$), which suggests that the association is not an artifact of the Gaussian assumption. We therefore regard the distance as an observational reliability diagnostic within the training distribution, and not as a calibrated out-of-distribution detector or a safety gate.

\subsubsection{Operating characteristics of the q40 rule and\texorpdfstring{\protect\newline}{} surrogate-only deployment}
\label{subsubsec:q40_results}

\begin{figure}[tbp]
    \centering
    \makebox[\textwidth][c]{\includegraphics[width=540pt,height=0.80\textheight,keepaspectratio]{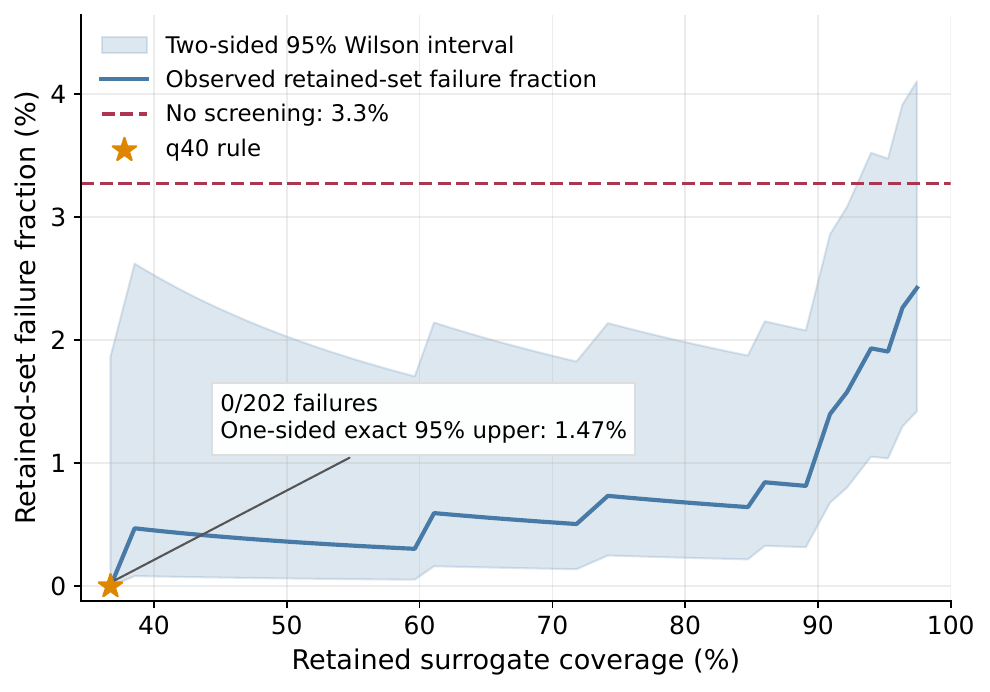}}
    \caption{Retained-set failure fraction (fraction of accepted scenarios with $R^2<0$) versus coverage for the SS-trained model (seed 0; 550 evaluation scenarios). Band: two-sided 95\% Wilson interval. Star: q40 rule (0/202) with its one-sided exact 95\% upper bound. Dashed line: no screening (18/550).}
    \label{fig:risk_coverage}
\end{figure}

\begin{figure}[tbp]
    \centering
    \makebox[\textwidth][c]{\includegraphics[width=540pt,height=0.80\textheight,keepaspectratio]{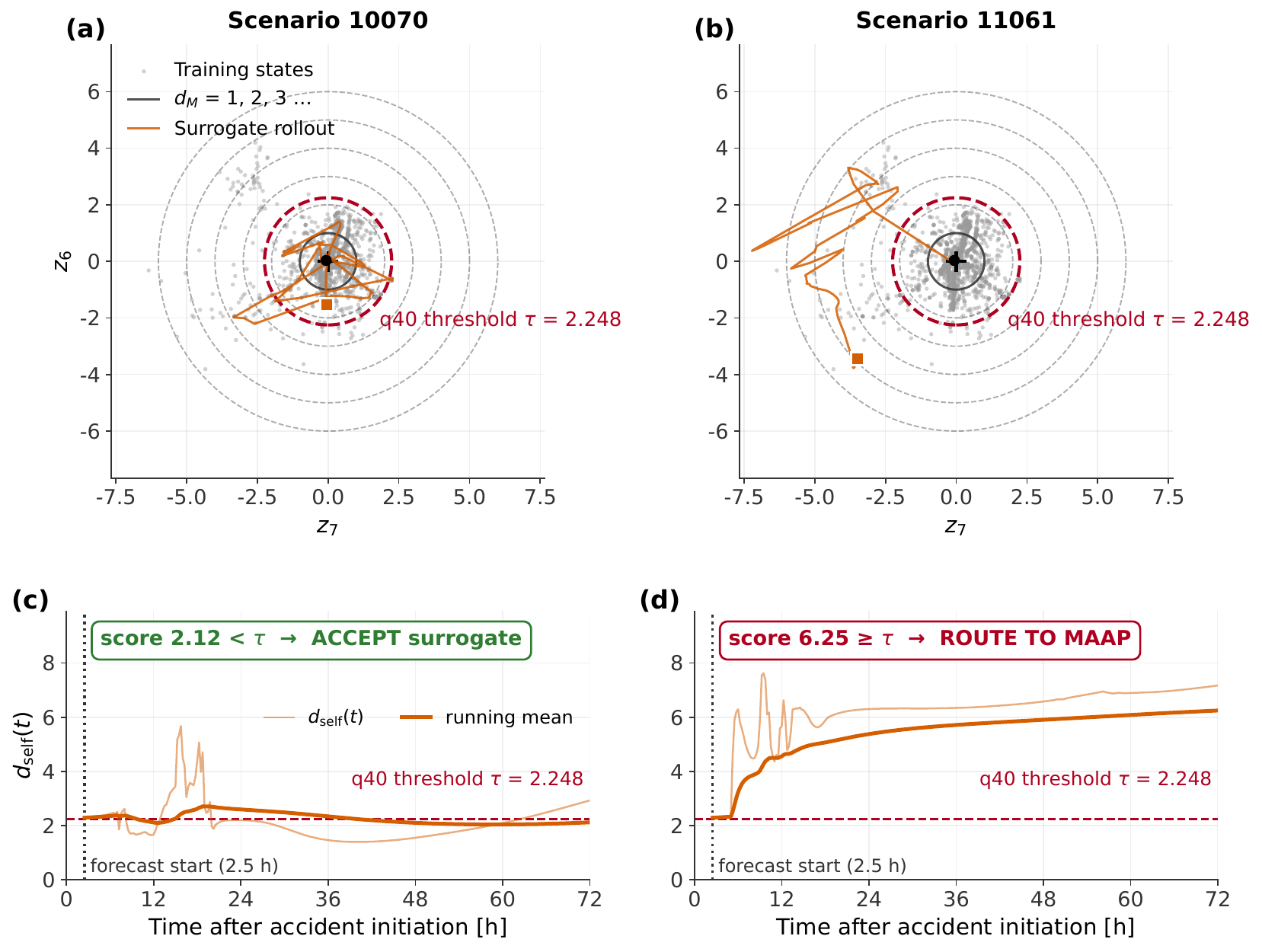}}
    \caption{Surrogate-only decision of the q40 rule (SS-trained model, seed 0) for two evaluation-half scenarios not used to set $\tau$: Scenario~10070 (typical) is accepted ($d_{\mathrm{self}} = 2.12$) and Scenario~11061 (lowest $R^2$) is routed to MAAP ($6.25$). (a,b) Whitened principal coordinates $z_7$ and $z_6$ of the training covariance, the two largest contributors to $d_{\mathrm{self}}$; in this 2-D projection of the 10-D space, the $\tau$ circle is a visual guide only. (c,d) Instantaneous $d_{\mathrm{self}}(t)$ and its running mean from the first forecast step (2.5~h); the decision uses the final time-averaged score. An animated version of this figure, showing how $d_{\mathrm{self}}(t)$ and its running mean
evolve along the rollout, is available at
\url{https://drive.google.com/file/d/1xca6Afa9FqPuc3c8_AUIMKcJARZ4S1BR/view}.}
    \label{fig:deployment}
\end{figure}

For the reference seed, the threshold $\tau = 2.2484$ accepts 202 of the 550 evaluation scenarios (36.7\%) and routes 348 to MAAP (Figure~\ref{fig:risk_coverage}). None of the accepted rollouts has $R^2<0$, whereas 18 scenarios (3.3\%) fail without screening; all 18 are routed. With zero failures among the 202 accepted rollouts, the nominal one-sided 95\% upper bound on the retained-set failure fraction is 1.47\%. The same holds for every training seed: no accepted rollout has $R^2<0$ (coverage $36.8 \pm 0.5\%$; bounds 1.45--1.49\%; thresholds 2.162--2.248; Table~\ref{tab:app_ss_seed_screen}). Among the operating points compared in Supplementary Material, Section S3, only q40 achieved this in all four seeds with a stable threshold, although the zero count depends on the calibration sample. The training-distance thresholds q90 and q92 accepted 5--8 failures per seed, and learn-then-test coverage ranged from 37.8\% to 90.0\% across seeds. The cost of the q40 rule is that about 63\% of the scenarios are routed to MAAP. Figure~\ref{fig:deployment} shows the decision for two scenarios.

Both scenarios come from the 550 evaluation-half scenarios not used to set $\tau$. Scenario~10070 is a typical accepted case ($R^2 = 0.939$, close to the median 0.938 of all 1100 scenarios; $d_{\mathrm{self}} = 2.12$, just below $\tau$). Scenario~11061 has the lowest $R^2$ of the SS-trained model ($-1.795$; $d_{\mathrm{self}} = 6.25$) and is routed to MAAP. The instantaneous $d_{\mathrm{self}}(t)$ may exceed $\tau$ transiently; the decision uses the final time-averaged score.

Accepted rollouts can still fail the CET checks. Over the four seeds, 26, 11, 8 and 26 accepted scenarios ($8.8 \pm 4.7\%$) violate at least one engineering criterion, against 21.3\% of the 550 evaluation scenarios without screening (reference seed). The self-distance ranks these violations only weakly (ROC-AUC 0.579). No accepted case misses a 1477-K crossing (per-variable errors: Supplementary Material, Section S9).
With MAAP substituted for the routed $63.2\%$, the hybrid output meets the 100-K peak-CET criterion in $97.0 \pm 1.6\%$ of the 550 evaluation scenarios (Eq.~\eqref{eq:hybrid_agreement}). This figure describes the combined workflow, including the MAAP runs for the routed scenarios. The routed fraction is the demand for MAAP confirmation runs (Section~\ref{subsec:operating_point}); net savings and runtime were not measured.

\subsubsection{Representative and worst-case scenarios}
\label{subsubsec:case_study}

To illustrate the distance--error relationship, we contrast a typical scenario of the SS-trained model with its worst scenario, both from the evaluation half. Scenario~10070 tracks the ground truth closely ($R^2 = 0.939$), whereas Scenario~11061, the lowest-$R^2$ case, fails ($R^2 = -1.795$, worse than a per-scenario mean predictor).
The worst case is selected by the lowest pooled per-scenario $R^2$. Scenario~1015, the largest-RMSE case of the TF-trained model ($R^2 = -5.63$), reaches $R^2 = -0.496$ with the SS-trained model. Supplementary Material, Section S10 traces both worst cases across the twelve configurations, in 4 of which Scenario~11061 has $R^2<0$.

\begin{figure}[tbp]
    \centering
    \makebox[\textwidth][c]{\includegraphics[width=540pt,height=0.80\textheight,keepaspectratio]{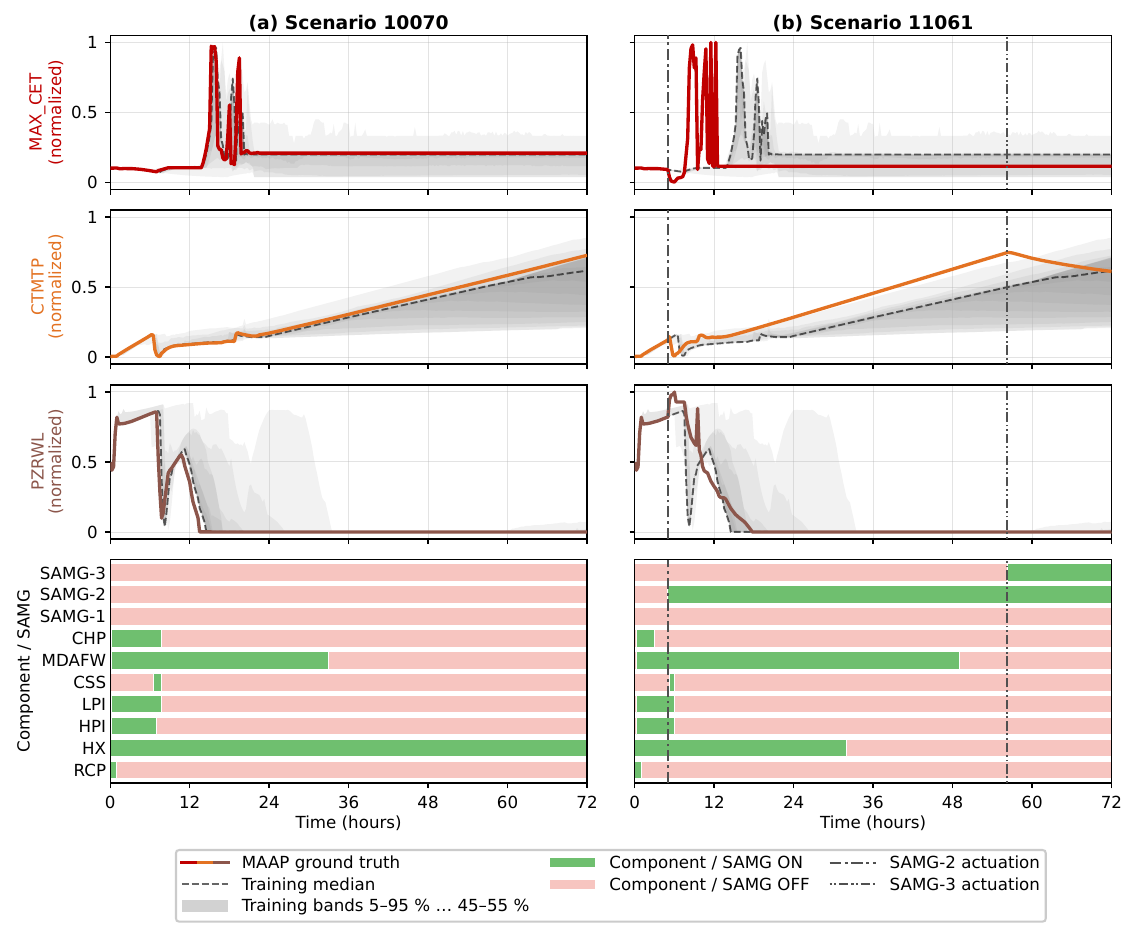}}
    \caption{MAAP ground-truth trajectories of three TH variables, component actions and SAMG states for Scenario~10070 (typical, accepted by the q40 rule) and Scenario~11061 (lowest $R^2$ of the SS-trained model, routed to MAAP). Gray: nested training percentile bands of the 9900 development scenarios (5--95\% to 45--55\%); dashed line: per-step training median.}
    \label{fig:density_plot}
\end{figure}

Figure~\ref{fig:density_plot} overlays three TH variables, safety-system actions, and SAMG intervention states of both scenarios on the percentile bands of the training distribution.

In Scenario~10070, no SAMG strategy is actuated, and MDAFW stops at 33.0~h. In Scenario~11061 (Figure~\ref{fig:density_plot}b), RCS depressurization (SAMG-2) starts early, at 5.0~h. HPI, LPI and containment spray stop at 6.0~h, the heat exchanger fails at 32.0~h and MDAFW stops at 49.0~h. RCS external injection (SAMG-3) starts at 56.25~h, and SAMG-1 is not actuated.
The q40 rule accepts Scenario~10070 ($d_{\mathrm{self}} = 2.12 < \tau$; peak-CET error 2.9~K) and routes Scenario~11061 ($d_{\mathrm{self}} = 6.25$) to MAAP. The routed case is a statistical failure ($R^2 = -1.795$), although its peak-CET error of 39.4~K lies within the 100-K tolerance. The statistical and engineering criteria can therefore disagree on the same scenario.

\subsection{Attention Score Analysis}
\label{subsec4-5}

The distance-to-training screen of Section~\ref{subsubsec:ood_screen} indicates \emph{which} rollouts are likely to be unreliable, but not \emph{why}. The attention weights give a post hoc view of which past time steps the model attends to. The analysis uses the archived TF-trained checkpoint (attention recorded along its autoregressive rollout) and Scenario~6736, that model's best-predicted scenario (lowest rollout RMSE; $R^2 = 0.998$), which was also the case study of the original analysis. Because the analysis was not recomputed for the SS-trained model, it describes what such a model attends to and does not explain why the screen flags a rollout. We examine the self-attention maps (Section~\ref{subsubsec:self_attention}) and the temporal pooling weights (Section~\ref{subsubsec:pooling_weights}) as a complement to the screen. Attention weights give only a partial account of a network's computation, and single-scenario plots are easily over-interpreted, so every general claim below is paired with an interval estimate over a 42-scenario sample (Supplementary Material, Section S11). The recency pattern of Scenario~6736, for example, holds across all 42 rollouts (Section~\ref{subsubsec:self_attention}).

\subsubsection{Inter-Temporal Dependencies via Self-Attention Maps}
\label{subsubsec:self_attention}

At every prediction step the self-attention layers assign a weight to each past position. Under causal masking these weights cover only earlier positions, so they indicate which earlier steps contribute to the current representation. Attention weights have been interpreted in this way in time-series modeling, for example in the Temporal Fusion Transformer for multi-horizon forecasting~\cite{Lim2021TFT} and in the RETAIN model for clinical sequences~\cite{choi2016retain}. Whether such weights faithfully explain a model's output is, however, still debated~\cite{jain2019attention, wiegreffe2019attention}, and we therefore present the maps below as descriptive
evidence and not as proof of a mechanism.

We probe the stack through the attention of the \emph{final} query over its
window. Because the prediction head uses the pooled mixture of all $k$ final-layer states (Eq.~\eqref{eq:context_vector}) and not only the last state, this row describes the dependencies inside the stack but not the full prediction. The pooling weights of Section~\ref{subsubsec:pooling_weights} describe the full prediction more directly. As the model is multi-head (Eq.~\eqref{eq:multihead}), we
report the average of this row over the $h$ heads of the final decoder layer (no averaging is performed across layers), giving
\begin{equation}
\label{eq:attention_last_row}
\mathbf{a}_{k} = \frac{1}{h}\sum_{j=1}^{h}
\mathrm{softmax}\!\left(\frac{\mathbf{q}_k^{(j)}\,{\mathbf{K}^{(j)}}^{\!\top}}
{\sqrt{d_k}}\right) = [a_{k,1},a_{k,2},\ldots,a_{k,k}],
\quad \sum_{i=1}^{k} a_{k,i}=1,
\end{equation}
where $\mathbf{q}_k^{(j)}$ and $\mathbf{K}^{(j)}$ are the query and keys of head
$j$. Each $a_{k,i}$ is the normalized weight the model places on past step $i$
when forming the representation of the most recent state, and $\mathbf{a}_k$ is
the vector analyzed throughout this section.

\begin{figure}[tbp]
    \centering
    \makebox[\textwidth][c]{\includegraphics[width=540pt,height=0.80\textheight,keepaspectratio]{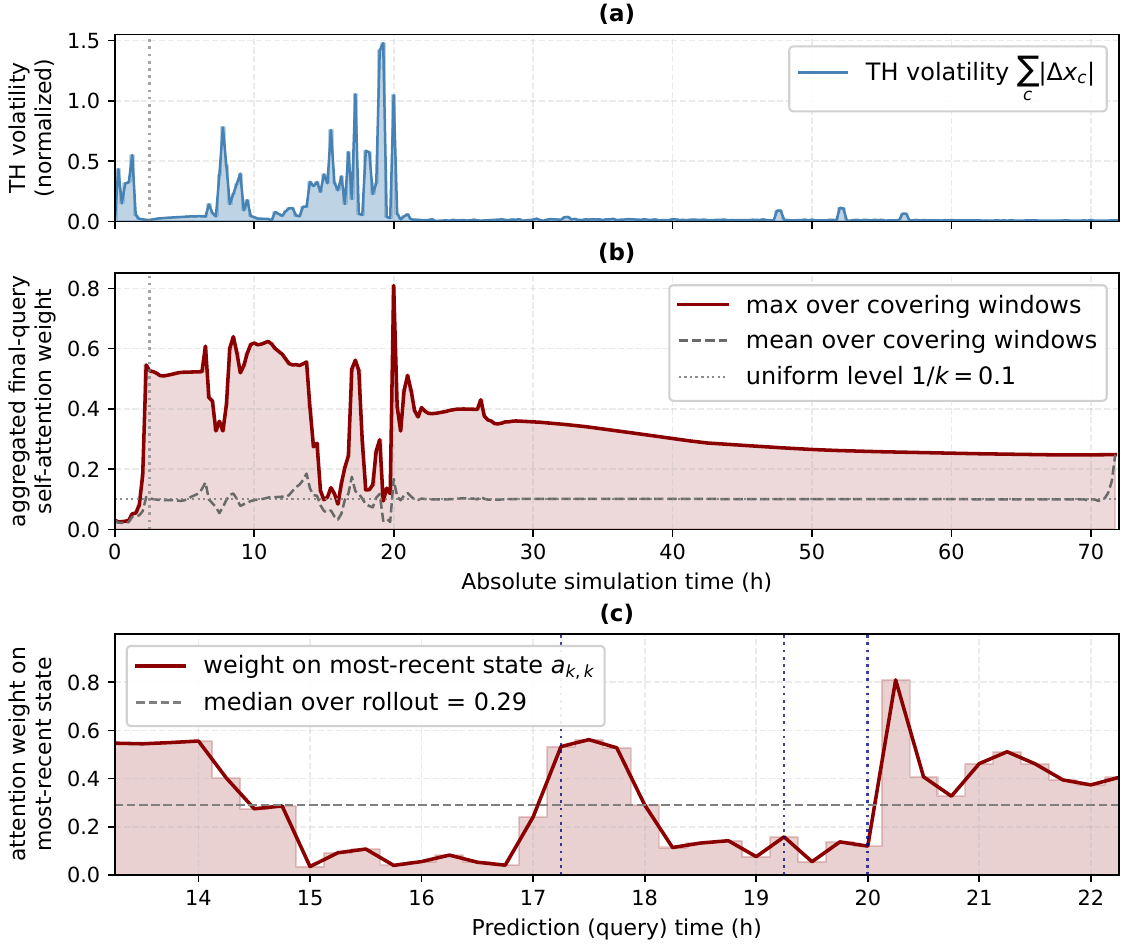}}
    \caption{Final-query self-attention of Scenario~6736 at the headline
    configuration (archived TF-trained checkpoint; attention recorded along its autoregressive rollout). (a)~Channel-summed one-step change of the TH state
    (volatility) over the full 72-hour transient. (b)~Largest and mean
    head-averaged weight each state receives across the rolling windows that
    contain it. (c)~Weight on the most-recent window position around the volatile phase;
    dotted lines mark the three strongest transients in the interval. Gray
    dotted vertical line in (a)--(b): forecast start ($t = 2.5$~h).}
    \label{fig:selfattn_kernel}
\end{figure}
As is common for autoregressive forecasters, the ABC-Transformer shows a recency bias and places most weight on the
steps just before the prediction. In the best-predicted scenario the weight vector peaks on the most-recent window position at $94.3\%$ of query steps. That position carries a weight of $0.29$ at the median step, and the center of mass lies $0.9$~h (\mbox{$3$--$4$} steps) in the past.
Figure~\ref{fig:selfattn_kernel} shows how this recency kernel behaves over the whole
transient. Panel~(b) records, for every state, the largest weight it
receives in any of the $k$ windows that contain it (the aggregation later applied to the pooling weights in Eq.~\eqref{eq:max_pooling}). Since that maximum
is attained while the state is the most recent one for $92\%$ of steps, the
curve reads as the salience of each state at the moment it was current. Its correspondence with the volatility of panel~(a) is direct. Through the volatile first $20$~h the peak salience runs at $0.5$--$0.6$ and reaches $0.8$ at the $20$-h burst, then relaxes to $0.25$--$0.34$ over the quiet remainder. The mean over windows, by contrast, stays pinned at the uniform level $1/k = 0.1$. The
kernel concentrates when the dynamics move and spreads when they do not, but
its mass never leaves the recent past. The pattern is population-wide rather than scenario-specific. Over a sample of $n = 42$ held-out scenarios the median
recency share is $95.0\%$, and every sampled scenario exceeds $90\%$ (Wilson
$95\%$ interval $[91.7, 99.9]\%$; Supplementary Table S11.1). The rolling window limits the attention span to $(k-1)\,\Delta t = 2.25$~h, so self-attention cannot reference an event older than one window, and any longer-range memory of the surrogate must be carried by the autoregressive state itself.

\subsubsection{Global Temporal Importance via Pooling Weights}
\label{subsubsec:pooling_weights}

Mean or max pooling would compress the window into a fixed vector without recording
which steps contributed. The ABC-Transformer instead pools the final-layer
states through a single learnable query
(Section~\ref{subsubsec:pooling}, Fig.~\ref{fig:overall_method}). This design follows the multi-head set pooling of Lee et al.~\cite{lee2019set}, reduced here to one query over the temporal axis. Unlike the self-attention rows above, these weights
enter the prediction head directly, which makes them a more suitable, although still post hoc, basis for physical interpretation.

Because the rolling window covers each physical time step many times, we summarize
the overlapping weights at step $t$ by their maximum across the windows
$\mathcal{W}_t$ that contain it,
\begin{equation}
\label{eq:max_pooling}
\bar{\alpha}_t^{\,\mathrm{max}} = \max_{w \in \mathcal{W}_t}\alpha_t^{(w)} .
\end{equation}
The maximum identifies steps that received a high weight in at least one window and
preserves the sharp event-level peaks that an average would smooth out.

Across the sampled scenarios (TF-trained checkpoint), the pooling weights tend to follow the dynamically
important phases of the transient. The per-scenario rank correlation between the state volatility and the pooled profile of
Eq.~\eqref{eq:max_pooling} is positive in $38$ of the $42$ sampled scenarios
($90.5\%$, Wilson $95\%$ interval $[78.1, 96.1]\%$). Its median is $+0.27$
(bootstrap $95\%$ CI $[0.23, 0.38]$; sign test $p = 5.7\times10^{-8}$;
Supplementary Material, Section S11). Whereas the self-attention rows are recency-dominated, the pooling stage is where event structure becomes legible. This matters because an analyst screening accident sequences
needs to see \emph{which} transitions the surrogate responded to, not only a
single error number.

\section{Conclusions}
\label{sec5}

This study introduced the Accident Binary--Continuous Transformer (ABC-Transformer), a decoder-only surrogate of the MAAP code for the OPR1000 total loss of component cooling water (TLOCCW) sequence. More importantly for reliability engineering, it characterized \emph{when} the model's autoregressive (AR) forecasts can be trusted. The architecture couples masked causal self-attention, which captures long-range temporal dependencies without recurrence, with a temporal attention pooling head. This head aggregates the entire observation window through a single learnable query rather than the final hidden state alone. Together these elements were designed to mitigate the vanishing-gradient and progressive error-accumulation behavior that constrains recurrent baselines on the long, event-driven trajectories typical of severe-accident progression.

Under one-step teacher forcing on a common dataset, the archived TF-trained ABC-Transformer attained a lower RMSE than all five baselines (CNN and recurrent models) at every resolution evaluated in the earlier study. At $\Delta t = 15$~min, it reduced MAE and RMSE by 35.3\% and 54.8\% against the strongest per-metric baselines, although at $\Delta t = 60$~min its MAE was marginally higher than that of the CNN hybrids. This single-run comparison is reported for protocol comparability only, because one-step accuracy does not carry over to the AR rolling-window regime in which the surrogate is used.

The main contribution is a distribution-level characterization of AR reliability over the 1100 held-out scenarios. Typical scenarios are reproduced well (SS-trained reference model: median per-scenario $R^2 = 0.938$). The lower macro-average ($0.836$) comes from a quantified failure tail ($3.2\%$ of scenarios below $R^2 = 0$; $3.8 \pm 1.0\%$ over four training seeds), not from degraded typical performance. AR error accumulates with rollout length, as expected from exposure bias. Scheduled-sampling training improved the aggregate error and the failure tail relative to the TF-trained model (macro-average $0.604 \to 0.836$; failure fraction $9.7\% \to 3.2\%$), which shows that the accumulation can be partly reduced through training. The self-distance of the surrogate's own rollout ranks the SS-trained failures with ROC-AUC $0.895$ (reference seed); it is available once the rollout is complete and before comparison with the MAAP reference. The q40 rule accepts $36.8 \pm 0.5\%$ of the evaluation scenarios with no accepted rollout below $R^2 = 0$ in any seed (nominal one-sided 95\% bounds $1.45$--$1.49\%$) and routes the rest to MAAP. The retrospective distance of the true trajectory is rank-correlated with AR error (Spearman $\rho = 0.70$, TF-trained model) but serves only as a diagnostic. No configuration dominated the median, the macro-average, and the failure fraction simultaneously. The headline configuration ($\Delta t = 15$~min, $k = 10$) was selected by the failure-first rule of Section~\ref{subsec4-2}.

Finally, an attention analysis of the archived TF-trained checkpoint, checked over a 42-scenario sample, indicated that the self-attention layers act mainly as an event-adaptive recency kernel. Event structure is more visible in the temporal pooling weights, which are positively correlated with the volatility of the transient in $38$ of $42$ scenarios. The TF-trained model's largest-RMSE scenario (1015) is among the few scenarios whose pooling profile gives no weight to the actuation that drives it out of the training band. Given the ongoing debate on whether attention is a faithful explanation, we present these observations as descriptive evidence rather than as a verified causal account.

The findings are limited in several respects, which determine how the model can be used.
\emph{Statistical scope.} Training-seed variation was assessed with four SS-trained models at the headline configuration only. The twelve-configuration grid and the archived TF-trained model are single runs (Supplementary Table S8.5 reports seeds retrained with the TF training pipeline). The reported intervals bound scenario-sampling uncertainty only. The configuration rule and the q40 level were chosen on the same 1100 test scenarios, so the q40 bounds are nominal and model-wise. Confirmation on newly generated scenarios is left for future work.
\emph{Physical scope.} Even in normalized terms, the errors on the fast safety-critical channels are large. For the TF-trained model, they reach $8.1\%$ of operating range on the steam-generator pressure and $4.8$--$6.6\%$ on the RCS loop and core-exit temperatures (Supplementary Material, Section S9). A signed-error analysis shows that the model under-predicts the escalating channels on the majority of rollout steps, so its pointwise errors lean non-conservative. For the SS-trained models, $8.8\,{\pm}\,4.7\%$ of q40-accepted scenarios (four seeds) violate at least one CET engineering criterion (Section~\ref{subsubsec:q40_results}). We therefore make no margin-to-setpoint statement and no conservative-bias guarantee. Consistent with the normalized-error convention (Eq.~\eqref{eq:normalization}), we report no physical-unit error bound beyond the CET study tolerances of Section~\ref{subsec:metrics}.
\emph{Task scope.} The binary safety-system and SAMG channels are supplied as known conditioning inputs. The problem solved is therefore conditional trajectory emulation given the realized actuation timeline extracted from the MAAP event log (sampled failure and SAMG-initiation times plus setpoint-triggered actuations). It is not autonomous prediction of operator or automatic actions. The predicted quantities are emulated MAAP outputs, not necessarily instrument-survivable plant signals during the degraded-core phase.
\emph{Generalization scope.} Training and evaluation were confined to a single accident (TLOCCW), a single plant (OPR1000), and a single code (MAAP), without MELCOR, plant-data, or integral-test cross-validation. Performance is unknown for other accident types, such as LOCA, steam generator tube rupture (SGTR), and station blackout (SBO). Because the distance-to-training analysis itself shows that accuracy degrades away from the training manifold, we claim no cross-accident, cross-plant, or cross-code generalization.
\emph{Screen scope.} $d_{\mathrm{self}}$ measures the distance of the predicted states from the development data. It is available before comparison with the MAAP reference but is not a calibrated failure probability, and a small value cannot rule out an error. The retrospective distance $d_{\mathrm{retro}}$ needs the MAAP trajectory and is a diagnostic only. The q40 level was selected by comparing operating points on the same scenarios, so it is specific to this model and scenario set.
\emph{Decision-channel accuracy and exposure bias.} The surrogate is least accurate on the SAMG-relevant variables (decision-channel median $R^2 = 0.61$, versus $0.90$ over all ten channels on the same per-scenario basis; TF-trained model, Supplementary Material, Section S9). Autoregressive error still grows with rollout length. Scheduled sampling reduces the aggregate error ($-38\%$ MAE, $-33\%$ RMSE) and more than halves the failure fraction (Section~\ref{subsubsec:training_seeds}), at the cost of a slightly smaller near-perfect fraction. Its seed-to-seed variation has been assessed at the headline configuration only.

Within these limits, the ABC-Transformer is best suited as an offline accelerator of MAAP-class analysis for scenario screening, sensitivity studies, and SAMG-strategy pre-screening. In this use, the q40 rule sends the cases that need confirmation to a full-code run. It is not a real-time decision-support tool, nor, given its non-conservative pointwise errors on the safety-critical channels, an instrument for establishing safety margins or qualifying SAMG decisions.

These limitations point to several directions for future work. Before any of them is pursued, the complete procedure (model, score, q40 rule, failure definitions) should be frozen and tested on newly generated scenarios, and the q40 level should be re-examined for other models and scenario sets. First, the observational distance-to-training screen could be turned into a calibrated, step-wise confidence bound by coupling it with conformal prediction~\cite{angelopoulos2023conformal, Yuan2026Conformal} or ensemble-based uncertainty quantification~\cite{lakshminarayanan2017ensembles, gal2016dropout, Basora2025UQBenchmark}. The conformal route would convert the present flag into a guaranteed-coverage reliability estimate. Second, a variable-step rollout scheme (multi-step increments in quiescent phases, single steps across rapid transients) could curtail the number of AR iterations and the associated error accumulation. Third, the gated variable-selection mechanism of the Temporal Fusion Transformer~\citep{Lim2021TFT} could be adopted to route the binary actuation signals and the continuous TH variables through dedicated pathways. This could improve accuracy on the SAMG-relevant channels that the present model handles least well. Fourth, because its inference is fast, the surrogate could serve as an environment simulator for reinforcement-learning-based operator-support and autonomous-control agents. Finally, instead of a separate model for each accident type, a unified surrogate could jointly learn LOCA, SGTR, and SBO sequences. Such a surrogate could improve robustness by sharing physical behavior across scenarios and widen the in-distribution envelope. Time-series foundation models pretrained on large and heterogeneous corpora~\cite{ansari2024chronos} may offer a route toward a transferable, multi-accident surrogate.

\section*{CRediT authorship contribution statement}
\textbf{Wonung Jeong:} Conceptualization, Methodology, Software, Validation, Formal analysis, Investigation, Visualization, Writing -- original draft.
\textbf{Sangam Khanal:} Conceptualization, Methodology, Software, Validation, Formal analysis, Investigation, Visualization, Writing -- original draft.
\textbf{Jeongik Lee:} Data curation.
\textbf{Sangseung Lee:} Conceptualization, Supervision, Writing -- review \& editing.
\textbf{Joongoo Jeon:} Conceptualization, Supervision, Project administration, Funding acquisition, Writing -- review \& editing.

\section*{Declaration of competing interest}
The authors declare that they have no known competing financial interests or personal relationships that could have appeared to influence the work reported in this paper.

\section*{Acknowledgements}
The authors would like to thank Semin Joo and Seokho Song (Korea Advanced Institute of Science and Technology) for their valuable contributions to this work. This work was supported by the K-Cloud project of Korea Hydro \& Nuclear Power Co., Ltd. (No.~2024-Tech-08). This work was also supported by the National Research Foundation of Korea (NRF) grant funded by the Korea government (MSIT) (No.~RS-2026-25541562).

\section*{Data availability}
The code (model, training, evaluation and screening scripts) and the trained checkpoints are available at \url{https://github.com/POSTECH-NINE/abc-transformer}. The MAAP simulation data that support the findings of this study are available from the corresponding author (J. Jeon) upon reasonable request.

\section*{Declaration of generative AI and AI-assisted technologies in the manuscript preparation process}
During the preparation of this work the authors used Claude (Anthropic) in order to check the manuscript against the journal formatting requirements and to improve the readability and organization of the text. After using this tool, the authors reviewed and edited the content as needed and take full responsibility for the content of the published article.

\input{appendix}

\bibliographystyle{elsarticle-num}
\bibliography{cas-refs}


\end{document}

%% file: Table06_reliability_distribution.tex
%

\begin{table}[htbp]
\centering
\caption{Per-scenario autoregressive reliability of the SS-trained model (seed 0) at the headline configuration
($\Delta t=15$~min, $k=10$, $n=1100$), using the per-scenario $R^2$ of Eq.~\eqref{eq:r2}.
Brackets denote 95\% confidence intervals over scenarios (bootstrap with $B = 5000$ for the median,
Wilson for the tail fractions); they bound scenario-sampling, not training-seed, uncertainty (seed variation: Table~\ref{tab:ss-seed-summary}).
The full per-configuration distribution is given in \ref{app:reliability}
(Table~\ref{tab:t2-ksens}). The TF-trained counterpart is Supplementary Table S8.3.}
\label{tab:t1-reliability-dist}
\small
\renewcommand{\arraystretch}{1.4}
\begin{tabular}{lr}
\toprule
Statistic (per-scenario $R^2$, autoregressive) & Value \\
\midrule
Median $R^2$                     & $0.938$~~[$0.932$,\,$0.945$] \\
Macro-average $R^2$              & $0.836$ \\
$10$th-percentile $R^2$          & $0.652$ \\
Fraction with $R^2<0$            & $3.2\%$~~[$2.3$,\,$4.4$]\% \\
Fraction with $R^2<0.5$          & $6.1\%$~~[$4.8$,\,$7.7$]\% \\
Fraction with $R^2\geq0.99$      & $18.9\%$ \\
Median RMSE / MAE (normalized)   & $0.037$ / $0.018$ \\
\bottomrule
\end{tabular}
\end{table}

%% file: Table07_distance_screen.tex
\begin{table}[H]
\centering
\caption{Spearman ($\rho$) and Pearson ($r$) correlations between distance-to-training measures and per-scenario autoregressive error of the SS-trained model (seed 0; $\Delta t = 15$~min, $k = 10$, $n = 1100$; observational, not causal). MD: retrospective Mahalanobis distance $d_{\mathrm{retro}}$ of the true (MAAP) trajectory (Eq.~\eqref{eq:mahalanobis}); $d_{\mathrm{self}}$: self-distance of the predicted trajectory (Eq.~\eqref{eq:self_distance}); PBE (pred.): percentile-band exceedance of the predicted trajectory. The TF-trained counterpart is Supplementary Table S2.1.}
\label{tab:t4-ood}
\small
\setlength{\tabcolsep}{5pt}
\renewcommand{\arraystretch}{1.2}
\begin{tabular}{llrrrr}
\toprule
Distance measure & Error metric & $\rho$ & $p_\rho$ & $r$ & $p_r$ \\
\midrule
MD (mean) & RMSE & $+0.649$ & $1.1\times10^{-132}$ & $+0.503$ & $1.5\times10^{-71}$ \\
MD (mean) & MAE & $+0.685$ & $3.6\times10^{-153}$ & $+0.531$ & $4.8\times10^{-81}$ \\
MD (mean) & $R^2$ & $-0.616$ & $4.9\times10^{-116}$ & -- & -- \\
MD (max) & RMSE & $+0.486$ & $2.0\times10^{-66}$ & $+0.193$ & $1.1\times10^{-10}$ \\
PBE (pred.) & RMSE & $-0.151$ & $4.7\times10^{-7}$ & $+0.003$ & $9.2\times10^{-1}$ \\
$d_{\mathrm{self}}$ & RMSE & $+0.576$ & $2.9\times10^{-98}$ & $+0.556$ & $3.1\times10^{-90}$ \\
\bottomrule
\end{tabular}
\end{table}

%% file: appendix.tex
\appendix
\setcounter{table}{0}\setcounter{figure}{0}
\makeatletter
\@addtoreset{table}{section}\@addtoreset{figure}{section}
\makeatother

\section{Hyperparameter tuning}
\label{app:hyperparams}

The base configuration was fixed in advance. The one-factor-at-a-time runs, including the configuration combining the per-factor optima (Table~\ref{tab:hp-tuning}, row (h)), quantify the sensitivity around it rather than select a configuration (Supplementary Material, Section S12). All main-text experiments (resolution and lookback grid, scheduled-sampling training and screening) use this base architecture. Table~\ref{tab:hp-tuning} presents the hyperparameter tuning results of the ABC-Transformer. Starting from the base configuration, we individually varied the model dimension, dropout rate, number of attention heads, number of decoder layers, learning rate, and lookback length. The train, validation, and test MSE are one-step teacher-forcing errors (in units of $10^{-4}$), whereas the AR test MSE is the autoregressive multi-step error (in units of $10^{-2}$). The archived TF-trained model uses the base architecture ($d_{\text{model}} = 64$, $h = 4$ attention heads, $N = 8$ decoder layers, $\eta = 10^{-3}$); the SS-trained models use the same architecture with the training settings of Table~\ref{tab:ss-config}. The tuning runs were trained and validated with one-step teacher forcing, whose error is only weakly sensitive to the lookback length (Supplementary Material, Section S5). The lookback deployed in the main text was selected separately under autoregressive rollout, where $k = 10$ minimizes the failure fraction (Section~\ref{subsec4-2}).

\begin{table}[H]
\centering
\footnotesize
\setlength{\tabcolsep}{3.5pt}
\sisetup{detect-weight=true}
\begin{threeparttable}
\caption{Hyperparameter tuning results of the ABC-Transformer.}
\label{tab:hp-tuning}
\begin{tabular}{cccccc S[table-format=1.3] S[table-format=1.3] S[table-format=1.3] S[table-format=1.3]}
\toprule
$d_{\text{model}}$ & $p$ & $h$ & $N$ & $\eta$ & $k$
& {Train MSE\tnote{a}} & {Val MSE$\downarrow$\tnote{a}} & {Test MSE\tnote{a,c}\hphantom{\textsuperscript{a,c}}} & {AR Test MSE$\downarrow$\tnote{b,d}\hphantom{\textsuperscript{b,d}}} \\
\midrule
256\tnote{e} & 0.2  & 4 & 8 & 1e{-}3 & 30 & 2.357 & 2.179 & 2.109 & \bfseries 0.611 \\
64  & 0.2  & 4 & 8 & 1e{-}4 & 30 & 3.223 & 2.664 & 2.622 & 1.085 \\
128 & 0.2  & 4 & 8 & 1e{-}3 & 30 & 2.227 & 2.058 & 1.961 & 1.266 \\
64  & 0.2  & 8 & 8 & 1e{-}3 & 30 & 2.226 & 1.985 & 1.956 & 1.540 \\
64  & 0.2  & 4 & 8 & 1e{-}3 & 15 & 2.312 & 2.092 & 1.985 & 1.595 \\
64  & 0.2  & 4 & 8 & 3e{-}4 & 30 & 2.441 & 2.146 & 2.094 & 1.678 \\
64\tnote{f}  & 0.2  & 4 & 8 & 1e{-}3 & 30 & 2.317 & 2.059 & 2.032 & 1.871 \\
64  & 0.01 & 4 & 8 & 1e{-}3 & 30 & 1.681 & 1.681 & 1.668 & 2.055 \\
64  & 0.2  & 4 & 8 & 1e{-}3 & 60 & 2.251 & 2.136 & 2.045 & 2.063 \\
64  & 0.2  & 2 & 8 & 1e{-}3 & 30 & 2.455 & 2.140 & 2.099 & 2.086 \\
64  & 0.2  & 4 & 4 & 1e{-}3 & 30 & 2.544 & 2.470 & 2.207 & 2.260 \\
32  & 0.2  & 4 & 8 & 1e{-}3 & 30 & 2.616 & 2.335 & 2.238 & 2.304 \\
64  & 0.2  & 4 & 8 & 1e{-}3 & 45 & 2.309 & 2.112 & 2.077 & 2.386 \\
64  & 0.05 & 4 & 8 & 1e{-}3 & 30 & 1.884 & 1.778 & 1.757 & 2.547 \\
64  & 0.1  & 4 & 8 & 1e{-}3 & 30 & 2.061 & 1.900 & 1.863 & 2.570 \\
64  & 0.2  & 4 & 2 & 1e{-}3 & 30 & 3.162 & 2.796 & 2.648 & 3.534 \\
64  & 0.3  & 4 & 8 & 1e{-}3 & 30 & 2.569 & 2.244 & 2.195 & 3.604 \\
\bfseries 64\tnote{g} & \bfseries 0.0 & \bfseries 4 & \bfseries 8 & \bfseries 1e{-}3 & \bfseries 30 & \bfseries 1.469 & \bfseries 1.613 & \bfseries 1.614 & 3.810 \\
\midrule
128\tnote{h} & 0.0  & 8 & 8 & 1e{-}3 & 30 & 1.509 & 1.771 & 1.692 & 5.554 \\
\bottomrule
\end{tabular}
\begin{tablenotes}[flushleft]
\footnotesize
\item[] \textbf{Abbreviations:}
\item[] $d_{\text{model}}$ -- model dimension
\item[] $p$ -- dropout rate
\item[] $h$ -- number of attention heads
\item[] $N$ -- number of decoder layers
\item[] $\eta$ -- learning rate
\item[] $k$ -- lookback length
\item[a] Reported in units of $10^{-4}$.
\item[b] Reported in units of $10^{-2}$.
\item[c] One-step (teacher-forcing) test MSE.
\item[d] Autoregressive multi-step test MSE.
\item[e] Best autoregressive configuration.
\item[f] Base configuration.
\item[g] Best one-step configuration.
\item[h] Combined configuration.
\end{tablenotes}
\end{threeparttable}
\end{table}

\begin{table}[H]
\centering
\caption{Scheduled-sampling training configuration of the SS-trained models (seeds 0, 42, 1, 2). The architecture is that of the base configuration of Table~\ref{tab:hp-tuning}; batch size, epoch limit and early-stopping patience differ from those of the archived TF-trained model.}
\label{tab:ss-config}
\small
\renewcommand{\arraystretch}{1.3}
\begin{tabularx}{\linewidth}{l X}
\toprule
Setting & Value \\
\midrule
Replacement probability $p$ & linearly annealed $0 \to 0.5$ \\
Replaced input & latest continuous row (detached prediction) \\
Optimizer & AdamW, lr $10^{-3}$ \\
Scheduler & StepLR (step $5$, $\gamma=0.2$) \\
Batch size / Epochs & $256$ / $25$ \\
TF validation MSE & $1.97\times10^{-4}$ (baseline $2.1\times10^{-4}$) \\
Dropout & $0.1$ \\
Early stopping & patience $6$ (teacher-forced validation MSE) \\
Seeds & $0$ (reference), $42$, $1$, $2$ \\
\bottomrule
\end{tabularx}
\end{table}

\section{Reliability distribution across all configurations}
\label{app:reliability}
This appendix reports the full per-scenario reliability distribution across all tested configurations.

\paragraph{Lookback effect under rollout.}
For the TF-trained model, Supplementary Table S8.4 ranks the lookback settings by three summaries that do not agree. By median per-scenario $R^2$, $k=30$ is highest at $\Delta t \leq 30$~min but is overtaken by $k=10$ at $\Delta t = 60$~min. The macro-average reverses this, ranking $k=30$ lowest at every resolution, and at $\Delta t = 15$~min the failure fraction favors $k=10$. For the SS-trained model (Table~\ref{tab:t2-ksens}), the failure fraction favors $k=10$ at every resolution, whereas the median favors $k=30$ only at $\Delta t = 15$ and $30$~min. No single configuration is therefore universally optimal, and the headline configuration follows the selection rule of Section~\ref{subsec4-2}.
\input{TableB1_lookback_sensitivity}

\clearpage
\section{Training-seed variation and seed-wise q40 screening}
\label{app:seeds}

\input{TableC1_seed_summary}

\input{TableC2_q40_by_seed}

Table~\ref{tab:ss-seed-summary} summarizes the four SS-trained models on the same 1100 test scenarios (Section~\ref{subsubsec:training_seeds}). The training seed changes parameter initialization, batch order and SS replacement draws, with the train/validation window split fixed. The standard deviations describe variation between trained models. Pooling their predictions would not create 4400 independent scenarios. Table~\ref{tab:app_ss_seed_screen} gives their q40 screening (Section~\ref{subsubsec:q40_results}).

%% file: TableB1_lookback_sensitivity.tex
\begin{table}[htbp]
\centering
\caption{Full per-configuration distribution of per-scenario autoregressive $R^2$ (Eq.~\eqref{eq:r2}) of the SS-trained model (seed 0)
over the test scenarios. Table~\ref{tab:t1-reliability-dist} gives the headline-configuration summary.
The headline configuration ($k=10$, $\Delta t=15$~min; in bold) minimizes the $R^2<0$ fraction within $\Delta t \leq 15$~min (Section~\ref{subsec4-2}). All errors are
normalized. All configurations use $n=1100$ scenarios except $k=30$ at $\Delta t=60$~min ($n=1098$). The TF-trained counterpart is Supplementary Table S8.4.}
\label{tab:t2-ksens}
\small
\renewcommand{\arraystretch}{1.3}
\begin{adjustbox}{max width=\linewidth,center}
\begin{tabular}{llrrrrrr}
\toprule
$\Delta t$ & $k$ & MAE\,med & RMSE\,med & $R^2$ mean & $R^2$ med & $R^2$ p10 & \%$R^2{<}0$ \\
\midrule
 & $3$ & 0.0601 & 0.0971 & \phantom{-}0.459 & 0.693 & $-0.008$ & 10.1 \\
 & $10$ & 0.0285 & 0.0496 & \phantom{-}0.726 & 0.919 & \phantom{-}0.302 & 5.8 \\
\multirow{-3}{*}{5~min} & $30$ & 0.0291 & 0.0486 & \phantom{-}0.482 & 0.892 & $-0.591$ & 16.2 \\
\midrule
 & $3$ & 0.0308 & 0.0547 & \phantom{-}0.518 & 0.885 & $-0.356$ & 15.8 \\
 & $\mathbf{10}$ & \textbf{0.0185} & \textbf{0.0365} & \phantom{-}\textbf{0.836} & \textbf{0.938} & \phantom{-}\textbf{0.652} & \textbf{3.2} \\
\multirow{-3}{*}{15~min} & $30$ & 0.0127 & 0.0224 & \phantom{-}0.841 & 0.962 & \phantom{-}0.654 & 3.5 \\
\midrule
 & $3$ & 0.0227 & 0.0405 & \phantom{-}0.812 & 0.917 & \phantom{-}0.591 & 3.1 \\
 & $10$ & 0.0114 & 0.0195 & \phantom{-}0.882 & 0.967 & \phantom{-}0.774 & 1.9 \\
\multirow{-3}{*}{30~min} & $30$ & 0.0090 & 0.0151 & \phantom{-}0.854 & 0.968 & \phantom{-}0.692 & 2.7 \\
\midrule
 & $3$ & 0.0128 & 0.0222 & \phantom{-}0.875 & 0.961 & \phantom{-}0.738 & 2.5 \\
 & $10$ & 0.0109 & 0.0188 & \phantom{-}0.894 & 0.963 & \phantom{-}0.756 & 1.4 \\
\multirow{-3}{*}{60~min} & $30$ & 0.0082 & 0.0135 & \phantom{-}0.496 & 0.939 & \phantom{-}0.619 & 4.1 \\
\bottomrule
\end{tabular}
\end{adjustbox}
\end{table}

%% file: TableC1_seed_summary.tex
\begin{table}[htbp]
\centering
\caption{Training-seed variation of the SS-trained model ($\Delta t = 15$~min, $k = 10$) on the same 1100 held-out test scenarios. MAE and RMSE (normalized) are scenario averages; failures are scenarios with per-scenario $R^2<0$. The last row gives the mean $\pm$ SD over four training seeds (SD: sample standard deviation).}
\label{tab:ss-seed-summary}
\small
\begin{tabular}{lrrrr}
\toprule
Seed & MAE & RMSE & Median $R^2$ & \makecell{Failures\\($R^2<0$)} \\
\midrule
0 (reference) & 0.0257 & 0.0468 & 0.938 & 35 \\
42            & 0.0270 & 0.0486 & 0.952 & 53 \\
1             & 0.0247 & 0.0450 & 0.948 & 30 \\
2             & 0.0262 & 0.0473 & 0.950 & 51 \\
\midrule
Mean $\pm$ SD & $0.0259 \pm 0.0010$ & $0.0469 \pm 0.0015$ & $0.947 \pm 0.006$ & \makecell{$42.2 \pm 11.5$\\($3.8 \pm 1.0\%$)} \\
\bottomrule
\end{tabular}
\end{table}

%% file: TableC2_q40_by_seed.tex
\begin{table}[htbp]
\centering
\caption{Seed-wise q40 screening of the SS-trained models ($\Delta t = 15$~min, $k = 10$) on the 550 evaluation scenarios. Threshold $\tau$, accepted ($n_a$) and routed counts, coverage $c = n_a/550$, accepted failures $k_f$ ($R^2<0$) and nominal model-wise bound $U_{0.95}$ (Eq.~\eqref{eq:cp_bound}). Eng.: accepted engineering violations. $G_{\mathrm{hybrid}}$: peak core-exit temperature error within 100~K (hybrid output).}
\label{tab:app_ss_seed_screen}
\small
\begin{tabular}{lrrrrrrrr}
\toprule
Seed & $\tau$ & $n_a$ & Routed & \makecell{$c$\\(\%)} & $k_f$ & \makecell{$U_{0.95}$\\(\%)} & Eng. & \makecell{$G_{\mathrm{hybrid}}$\\(\%)} \\
\midrule
0 (reference) & 2.2484 & 202 & 348 & 36.7 & 0 & 1.47 & 26 & 95.8 \\
42            & 2.1619 & 199 & 351 & 36.2 & 0 & 1.49 & 11 & 98.2 \\
1             & 2.2090 & 205 & 345 & 37.3 & 0 & 1.45 & 8  & 98.7 \\
2             & 2.2167 & 203 & 347 & 36.9 & 0 & 1.46 & 26 & 95.5 \\
\bottomrule
\end{tabular}
\end{table}